\documentclass[aps,twocolumn,prl,showpacs,amsmath,amssymb,longbibliography,showkeys,10pt]{revtex4-2}
\usepackage{graphicx}
\usepackage{epstopdf}
\usepackage[normalem]{ulem}
\usepackage{braket}
\usepackage{hyperref}
\usepackage[dvipsnames]{xcolor}
\usepackage{ulem}
\allowdisplaybreaks

\begin{document}

\title{Operating a contextual Stern--Gerlach apparatus}

\author{Th. K. Mavrogordatos}
\email{themis.mavrogordatos@fysik.su.se}
 \affiliation{Department of Physics, AlbaNova University Center, SE 106 91, Stockholm, Sweden}

\date{\today}

\begin{abstract}
We propose a contextual cavity/circuit QED analogue and extension of the Stern--Gerlach experiment, where the pseudo-spin of a two-state `atomic' transition plays the role of the ``spin'', while the resonant field driving the transition stands for the ``magnetic field''. A phase-sensitive continuous detection of the cavity field coupled to the induced `atomic' dipole affects the stability of the two distinct outcomes. The dressed states comprising the latter give their place to a self-consistent spontaneous dressed-state polarization as the driving strength is lowered. The associated evolution proves anew highly contextual, underpinned by a persistent production of coherent-state superpositions for a particular setting of the monitoring device. Finally, when bistability is absent, we employ the photoelectron `atomic' emission statistics as a diagnostic tool of the cavity field fluctuations.  
\end{abstract}


\maketitle

In a Stern--Gerlach apparatus, a quantum particle is forced into revealing its spin state by the inhomogeneous part of an external magnetic field. As the experiment pans out, a correlation is gradually established between the center-of-mass and spin state of the particle and, over a sufficient lapse of time, alternative center-of-mass positions can be readily discerned. A laser light intensity gradient perpendicular to an atomic beam achieved the said correlation in~\cite{Sleator1996}. With the advent of circuit QED~\cite{Blais2021, GarcíaRipoll2022, Minev2019}, the confinement of quantum trajectories to a meridian or the equator of the Bloch sphere was realized with a phase-sensitive operation scheme~\cite{Qubit2013, Weber2016}, following the extraction of a stochastic equation for the qubit in the dispersive regime of the Jaynes--Cummings (JC) model~\cite{Gambetta2008}. In parallel, paradigmatic cavity and circuit QED experiments have captured the characteristic JC nonlinearity in an explicitly open quantum system context~\cite{BreuerPetruccione} and in various regimes of operation displaying pronounced nonlinearity~\cite{Fink2008, Bishop2009, Kerckhoff11, Fink2017, Sett2024}.
\begin{figure}
\includegraphics[width=0.5\textwidth]{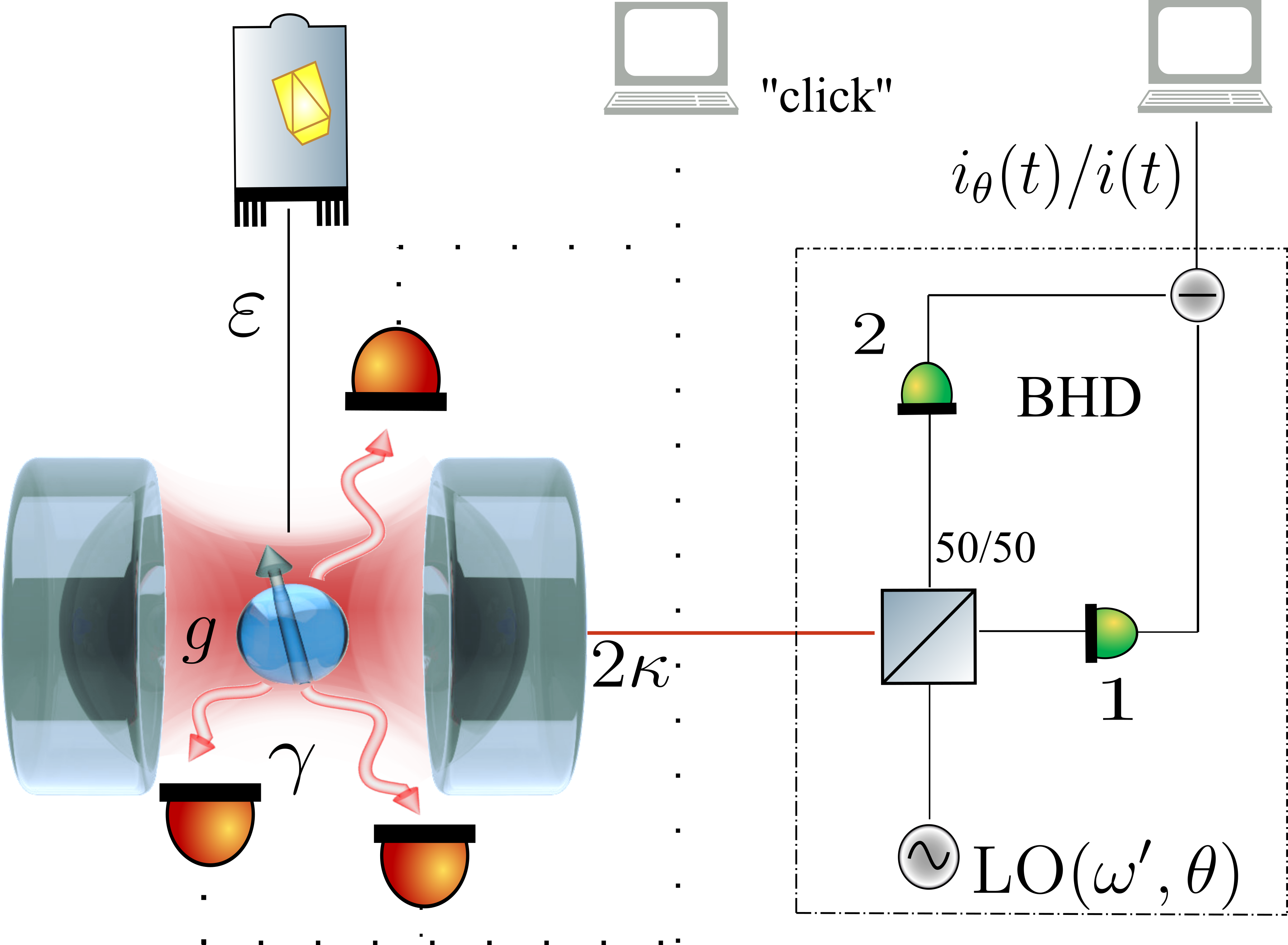}
\caption{{\it Schematic illustration of the Stern--Gerlach apparatus.} Fluorescence from a resonantly driven two-state `atom' is superimposed with a strong local oscillator (LO) in a balanced detection scheme. A cavity with photon loss rate $2\kappa$ collects forward radiation into a resonant mode of frequency $\omega$, and directs it out one of its mirrors. The measurement signal $i_{\theta}(t)$ (homodyne scheme, $\omega^{\prime}=\omega$ and fixed $\theta$) or $i(t)$ [heterodyne scheme, $\omega^{\prime}-\omega \gg \kappa$ and $\theta=(\omega-\omega^{\prime})t$] arises from the detection of two fields with a $\pi$ phase difference between their superpositions. Simultaneously, cavity-inhibited spontaneous emission is collected from detectors situated at the sides of the cavity. The conditional wavefunction evolves according to Eq.~\eqref{eq:SSE} between `atomic' collapses.}
\label{fig:FIG1}
\end{figure}

In this Letter, we are considering a resonant JC coupling between a two-state `atom' and the cavity field, where both degrees of freedom are kept in place. When the `atom' is in the dressed states of its interaction with an external coherent field it has an induced dipole moment. The phase of the radiated light depends on the phase of the induced dipole, which differs by $\pi$ between the two dressed states. Therefore, detection of dipole radiation in a phase sensitive manner will act as the analogue of measuring the particle position in the conventional Stern--Gerlach experiment~\cite{CarmichaelSG1994}. Thus far, such an analogue of the pivotal experiment has not been realized with a phase-sensitive continuous detection of the output field. This report reveals the operational consequences of quantum contextuality~\cite{Carmichael1999, CarmichaelBook2} inseparably attached to the above measurement strategy, and correlated with markedly different manifestations of the JC nonlinear excitation spectrum in the output field. 

The two-state `atom' is coherently driven by a field of amplitude $\varepsilon$ and frequency $\omega$, in resonance with the `atomic' transition and the cavity mode. In a frame rotating with the drive, we write the Lindblad master equation (ME) in the rotating wave approximation, 
\begin{equation}\label{eq:ME}
\frac{d\rho}{dt}=\frac{1}{i\hbar}[H, \rho] + \mathcal{L}[\sqrt{2\kappa}\,a]\rho  + \mathcal{L}[\sqrt{\gamma}\,\sigma_{-}]\rho 
\end{equation}
where $\mathcal{L}[\xi] \cdot \equiv \xi \cdot \xi^{\dagger}-\tfrac{1}{2}(\xi^{\dagger}\xi \cdot +\cdot \xi^{\dagger}\xi)$ and
\begin{equation}\label{eq:JCHam}
\begin{aligned}
H&=i\hbar g (a^{\dagger}\sigma_{-}-a\sigma_{+}) + \hbar \varepsilon (\sigma_{+}+\sigma_{-}). 
 \end{aligned}
\end{equation}
In the above Hamiltonian, $a^{\dagger}$ and $a$ are the creation and annihilation operators for the cavity field, $\sigma_{+}=|+\rangle\langle-|$ and $\sigma_{-}=\sigma_{+}^{\dagger}$ raise and lower the `atomic' states $|-\rangle$ and $|+\rangle$, respectively, while $g$ is the dipole coupling strength. The quadrature phase operators of the cavity field dissipated at rate $\kappa$, are defined as $\mathcal{A}_{\theta}\equiv\frac{1}{2} (a e^{-i\theta} + a^{\dagger}e^{i\theta})$. The cavity output is continuously measured via homodyne or heterodyne detection~\cite{Carmichael1993QTIII, Wiseman1993, CarmichaelBook2}, and the cavity-inhibited spontaneous emission from the `atom' at rate $\gamma$ is recorded via {\it direct} photoelectron detection~\citep{Glauber1963, Glauber1963II, KelleyKleiner1964, Saleh1978, Carmichael1993QTI, Carmichael1989}. The experimental setup is illustrated in Fig.~\ref{fig:FIG1}, extending the original proposal of~\citep{CarmichaelSG1994}.

\begin{figure*}
\includegraphics[width=\textwidth]{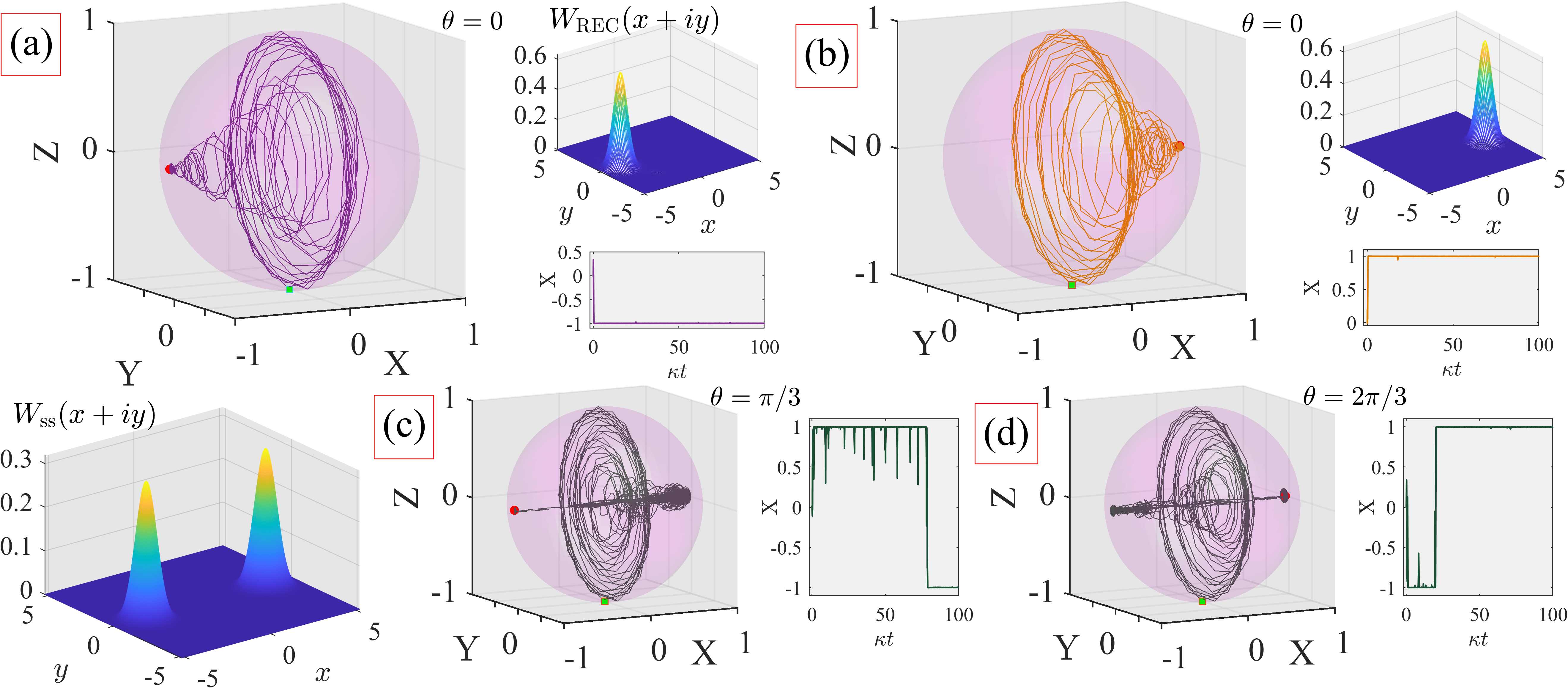}
\caption{{\it Contextual stability of stationary states.} Sample trajectories [solutions to Eq.~\eqref{eq:SSE}] in the Bloch-sphere representation $(X,Y,Z)$, with $X-iY=2\langle \psi_{\rm REC}(t)|\sigma_{-}|\psi_{\rm REC}(t)\rangle$ and $Z=\langle \psi_{\rm REC}(t)|\sigma_{z}|\psi_{\rm REC}(t)\rangle$, in the course of $100$ cavity lifetimes. The green square and the red circle mark the start and end of the trajectory, respectively. The LO phase is set to $\theta=0$ in frames {\bf (a, b)}, $\pi/3$ in {\bf (c)} and $2\pi/3$ in {\bf (d)}. The upper side panels in {\bf (a, b)} depict the conditioned Wigner function of the cavity field $W_{\rm REC}(x+iy)$ at $\kappa t=60$. The lower ones plot $X(t)$, as do the side panels of {\bf (c, d)}. The bottom left surface plot depicts the steady-state Wigner function of the resonant cavity mode, $W_{\rm ss}(x+iy)$. Operating parameters read: $\varepsilon/\kappa=300$, $g/\kappa=7$ and $\gamma/(2\kappa)=0$. In all simulations, the Fock-state basis was truncated at the $35$-th photon level and the JC system was initialized in $|0\rangle |-\rangle$.}
\label{fig:FIG2}
\end{figure*}

Let us for the moment neglect the coupling of fluorescence to the resonant cavity mode (set $g=0$) and focus on the evolution of a ``spin'' under the sole action of an effective ``magnetic'' field. In the analogue we are studying, the Stern--Gerlach ``spin'' states are the dressed states~\cite{Alsing1991}
\begin{equation}\label{eq:dressed}
\ket{\uparrow}\equiv (1/\sqrt{2})(|+\rangle + |-\rangle) \quad {\rm and} \quad \ket{\downarrow}\equiv (1/\sqrt{2})(|+\rangle - |-\rangle),
\end{equation}
with energies $+\hbar \varepsilon$ and $-\hbar\varepsilon$, respectively. The ``magnetic'' field, defined as $B \equiv (\varepsilon, 0 ,0)$, is taken along the direction of $\ket{\uparrow}$, $\ket{\downarrow}$ ($x$-axis) in the Bloch-sphere representation (with $|+\rangle$ and $|-\rangle$ aligned along the $z$-axis). If the two-state `atom' is prepared in the ground state $|-\rangle=(\ket{\uparrow}-\ket{\downarrow})/\sqrt{2}$, it evolves to $(e^{-i\varepsilon t}\ket{\uparrow}-e^{i\varepsilon t}\ket{\downarrow})/\sqrt{2}$, {\it i.e.} it exhibits a continuous Rabi oscillation with frequency $2\varepsilon$. To bring about an evolution to distinct states as registered by a Stern--Gerlach apparatus, we will use quantum trajectory theory~\cite{Barchielli1991, Dalibard1992, Molmer1996, Carmichael2013Ch4} to unravel the full ME~\eqref{eq:ME} via homodyne and heterodyne detection. Once we explore the complementary conditioned evolution for $\varepsilon/g \gg 1$, we will extend the relaxation to {\it distinct states} to the emergence of {\it distinct patterns} in the steady-state dynamics, shaped by the interplay of quantum coherence and measurement backaction~\cite{Hatridge2013}. The background is set by the resonant JC nonlinearity~\cite{Savage1988, Alsing1991, Kilin91, Alsing1992, Armen2009, Carmichael2015, Curtis2021}.  

We return to the open-quantum-system formulation. The un-normalized conditioned wavefunction for the two-state `atom' and cavity mode `molecule', obeys the stochastic Schr\"{o}dinger equation (SSE)~\cite{CarmichaelSG1994, Reiner2001}
\begin{equation}\label{eq:SSE}
\begin{aligned}
d|\psi_{\rm REC}(t)\rangle=&\big\{[-\kappa a^{\dagger}a -(\gamma/2) \sigma_{+}\sigma_{-}\\
&\quad- i\varepsilon(\sigma_{+}+\sigma_{-}) + g(a^{\dagger}\sigma_{-}-a\sigma_{+})]dt\\
&+\sqrt{2\kappa}\,e^{-i\theta}\,a\,dq_{\theta}(t)\big\}\,|\psi_{\rm REC}(t)\rangle,
\end{aligned}
\end{equation}
where
\begin{equation}\label{eq:dq}
dq_{\theta}(t)=\sqrt{8\kappa}\langle \psi_{\rm REC}(t)| \mathcal{A}_{\theta}|\psi_{\rm REC}(t)  \rangle\, dt + dW
\end{equation}
is the incremental charge (divided by an appropriate gain coefficient) released from the detector in the interval $t$ to $t+dt$. In the above expression, $dW$ is a real-valued Wiener increment (with zero mean and covariance $\overline{dW^2}=dt$) associated with the random emission of photoelectrons at an average rate dominated by the local oscillator (LO) photon flux (see Fig.~\ref{fig:FIG1}). The LO amplitude has a definite phase $\theta$, with $0 \leq \theta < \pi$, introducing a clear phase reference to the phase-dependent phenomenon we are studying: it pertains to both the field quadratures $\mathcal{A}_{\theta}$ and the phased emission of the induced dipole. 

Equation~\eqref{eq:dq} makes the measured signal following suitable post-processing. In particular, the infinite-bandwidth current derived from $dq_{\theta}(t)$ is filtered with a certain detection bandwidth $B$; the resulting photocurrent $i_{\theta}(t)$ obeys the equation~\cite{CarmichaelBook2}
\begin{equation}
di_{\theta}=-B(i_{\theta}\,dt - dq_{\theta}).
\end{equation}

The phases of the coupling of the `atom' to the coherent drive ($\varepsilon$) and cavity mode ($g$) have been selected so that the radiation impinging on the detector is either in phase or $\pi$ out of phase with the LO field {\it for $\theta=0$}, depending on the dressed ``spin'' state of the `atom'. Single realizations are generated via a quantum Monte Carlo algorithm solving Eq.~\eqref{eq:SSE} for the field initialized in the vacuum state and the `atom' in $|-\rangle$. Steady-state results are generated through a numerical diagonalization of the Liouvillian~\cite{Tan1999} corresponding the ME~\eqref{eq:ME}.

We first operate in the ``zero system size'' $\gamma/(2\kappa)=0$~\citep{Carmichael2015} and examine the limit $\varepsilon/g \gg 1$, in which the steady-steady solution of the ME~\eqref{eq:ME} is bimodal with the two coherent-state amplitudes placed along the real axis, and the `atomic' states are given by Eq.~\eqref{eq:dressed}. The coherent states comprising the statistical mixture of the cavity state have amplitudes $\alpha_{\rm ss}=\pm g/(2\kappa)$, as we can observe in Fig.~\ref{fig:FIG2}. In fact, we may anticipate this behaviour by rewriting Eq.~\eqref{eq:SSE} in terms of the dressed-state `atomic' operators
\begin{equation}
d_z=\ket{\uparrow}\bra{\uparrow} - \ket{\downarrow} \bra{\downarrow}, \quad d_{+}=\ket{\uparrow} \bra{ \downarrow}, \quad d_{-}=d_{+}^{\dagger}.
\end{equation}
Doing so we obtain
\begin{equation}\label{eq:SSEdressed}
\begin{aligned}
d|\psi_{\rm REC}(t)\rangle=&\big\{[-\kappa a^{\dagger}a - i\varepsilon\,d_z + (g/2)d_z(a^{\dagger}-a)\\
&+(g/2)(d_{+}-d_{-})(a^{\dagger}+a)]dt\\
&+\sqrt{2\kappa}\,e^{-i\theta}\,a \,dq_{\theta}(t)\big\}\,|\psi_{\rm REC}(t)\rangle.
\end{aligned}
\end{equation}
To remove the term $- i\varepsilon\,d_z$ we transform to the interaction picture; doing so, we observe that the dressed operators $d_{+}(t)$ and $d_{-}(t)$ oscillate at the high frequency $\varepsilon$ and yield a negligible contribution when averaged over a timescale $\sim 1/g$. We can therefore omit $(g/2)(d_{+}-d_{-})(a^{\dagger}+a)$ in comparison to $(g/2)d_z(a^{\dagger}-a)$ when effecting the familiar {\it secular approximation}~\cite{CohenTannoudji1977, CarmichaelBook2}, and seek stationary states of the stochastic evolution. The latter must satisfy
\begin{equation}
\begin{aligned}
\{[-\kappa a^{\dagger}a +\tfrac{g}{2}d_z(a^{\dagger}-a)]dt& + \sqrt{2\kappa}\,e^{-i\theta}\,a \,dq_{\theta}(t)\}|\psi_{\rm REC}(t)\rangle\\
 &= dE(t) |\psi_{\rm REC}(t)\rangle,
\end{aligned}
\end{equation} 
where $E(t)$ plays the role of the open-system {\it quasi}energy as a complex-valued stochastic process which explicitly encompasses the effect of continuous measurement backaction for different settings of $\theta$. We use a separable-state {\it ansatz} of the form $|\alpha\rangle \ket{\uparrow (\downarrow)}$, with $|\alpha \rangle$ a coherent state of the cavity field. Substituting yields
\begin{equation}\label{eq:ast}
\begin{aligned}
\{[-\kappa a^{\dagger}\alpha \pm \tfrac{g}{2}(a^{\dagger}-\alpha)]dt& + \sqrt{2\kappa}\,e^{-i\theta}\,\alpha \,dq_{\theta}(t)\}|\alpha\rangle \ket{\uparrow (\downarrow)}\\
 &= dE(t) |\alpha\rangle \ket{\uparrow (\downarrow)},
\end{aligned}
\end{equation}
which can be satisfied by choosing the states $|g/(2\kappa)\rangle \ket{\uparrow}$ or $|-g/(2\kappa)\rangle\ket{\downarrow}$. 

To assess whether the stationary coherent states with $\alpha=\pm g/(2\kappa)$ are stable against deviations induced by measurement backaction, we can evaluate the `deterministic' part of the {\it quasi}energy differential $dE$ ({\it i.e.}, excluding the shot-noise contribution) from Eq.~\eqref{eq:ast} as
\begin{equation}\label{eq:dE}
dE^{\prime}=[\mp g \alpha/2 + 4\kappa \alpha^2\cos^2\theta -4i\kappa \alpha^2 \cos\theta \sin\theta]\,dt,
\end{equation}
independent of the sign of $\alpha$. For $\theta=0$, $dE^{\prime}$ is real as are the shot-noise fluctuations. The `atomic' state relaxes to either $\ket{\uparrow}$ or $\ket{\downarrow}$ in a fraction of a cavity lifetime under the action of the term $\sqrt{2\kappa}\,a \,dW$ in Eq.~\eqref{eq:SSE} (the Wiener increment takes either positive of negative values with equal probability) and that choice is definite, as we can observe in Figs.~\ref{fig:FIG2}(a, b). Trajectories reach distinct attractors with either $X=+1$ or $X=-1$ (compare with Fig. 3 of~\citep{CarmichaelSG1994}). The stationary Wigner function~\cite{SchleichCh3, CarmichaelBook1} of the cavity field evinces a coherent state centred at $g/(2\kappa)$ and $-g/(2\kappa)$, respectively. This trend, however, is particular to ``zero system size'': in the presence of cavity-inhibited spontaneous emission, the above states have a finite lifetime equal on average to $4/\gamma$~\cite{Alsing1991}.

\begin{figure*}
\includegraphics[width=\textwidth]{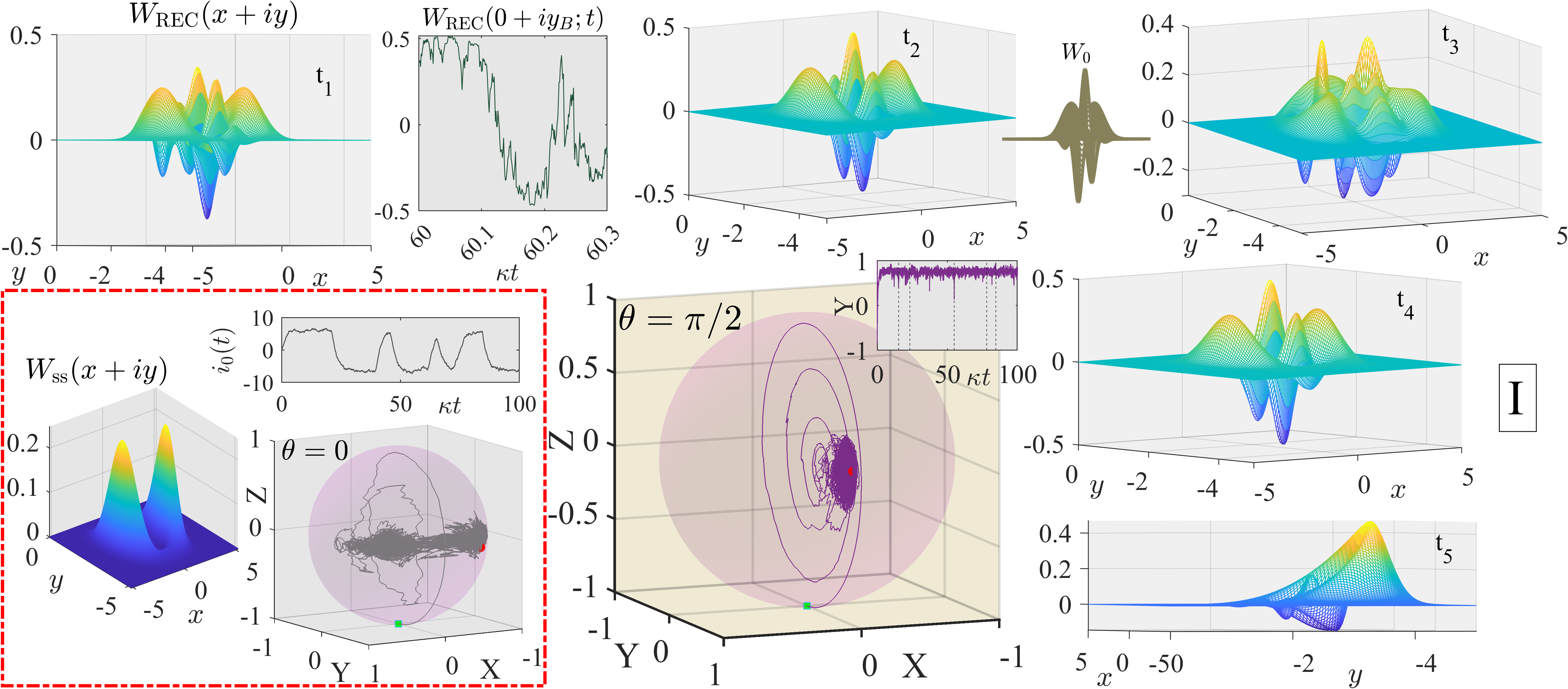}
\noindent\rule{18cm}{0.7pt}
\vspace{5mm}
\includegraphics[width=\textwidth]{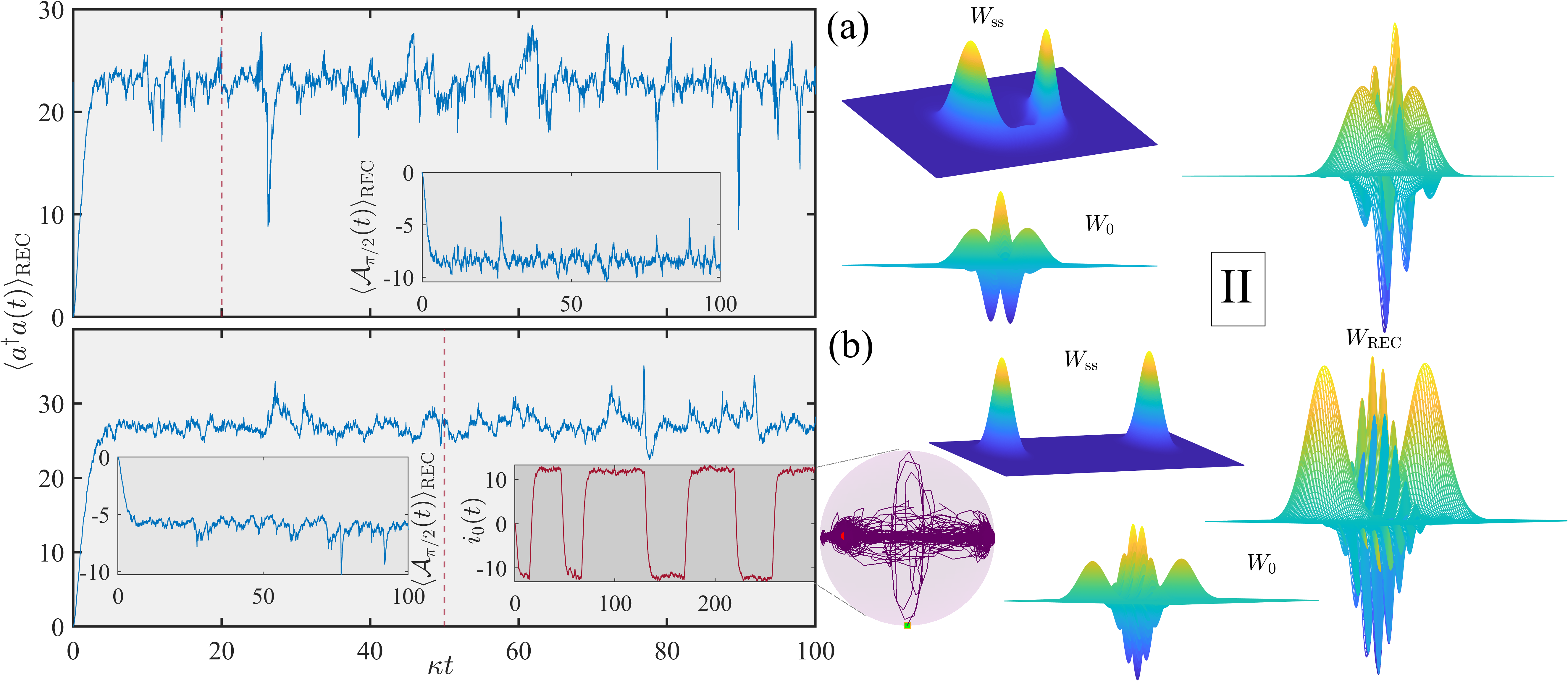}
\caption{{\it Spontaneous dressed-state polarization vs. persistence of coherent-state superpositions.} {\bf Panel I:} The central frame depicts a sample trajectory in the Bloch-sphere representation $(X,Y,Z)$, in a similar fashion to Fig.~\ref{fig:FIG2}, with the LO phase set to $\theta=\pi/2$. The surrounding conditioned Wigner functions $W_{\rm REC}(t)$ are extracted for the times $\kappa (t_1, t_2,\ldots,t_5)=15.20, 23.34, 54.89, 78.02$ and $84.57$, indicated by the dashed vertical lines in the corresponding trajectory for $Y(t)$. The plot between frames $t_1$ and $t_2$ depicts the conditioned Wigner function at the point $(0,y_B)$ of phase space (with $y_B=-2.86$), against $0.3$ cavity lifetimes starting from $\kappa t=60$. The schematic surface plot between frames $t_2$ and $t_3$ depicts the Wigner function $W_0$ corresponding to the (appropriately normalized) superposition state $|\alpha_{B+}\rangle + e^{i\phi} |\alpha_{B-}\rangle$ with $\phi=0$. In the framed group of panels we find another trajectory in the Bloch-sphere representation, generated for $\theta=0$, below the corresponding homodyne photocurrent. On the left we find the steady-state Wigner function of the resonant cavity mode, $W_{\rm ss}(x+iy)$. Operating parameters read: $\varepsilon/\kappa=30$, $g/\kappa=7$, $B/\kappa=0.5$ and $\gamma/(2\kappa)=0$. In all simulations, the Fock-state basis was truncated at the $35$-th photon level, and the JC system was initialized in $|0\rangle |-\rangle$. {\bf Panel II:} Individual quantum trajectories of the cavity photon number $\langle a^{\dagger}a(t) \rangle_{\rm REC} \equiv \langle \psi_{\rm REC}(t)|a^{\dagger}a|\psi_{\rm REC}(t)\rangle$ and the quadrature phase amplitude $\langle \mathcal{A}_{\pi/2}(t) \rangle_{\rm REC} \equiv \langle \psi_{\rm REC}(t)| \mathcal{A}_{\pi/2}|\psi_{\rm REC}(t)\rangle$ (insets) against $100$ cavity lifetimes for $\varepsilon/g=5.5$ in {\bf (a)} and $10$ in {\bf (b)}. The LO phase is fixed at $\pi/2$ in both (a, b), while the bistability threshold is located at $\varepsilon/g=g/(2\kappa)=5$. The right inset inside the main plot of (b) depicts the photocurrent $i_0(t)$ extracted from a different realization with the LO phase set to $0$, against $300$ cavity lifetimes, joined to the corresponding Bloch-sphere representation of the trajectory. Schematic surface plots on the right depict the corresponding steady-state Wigner functions $W_{\rm ss}$, Wigner functions $W_0$ of the (normalized) cat states $|\alpha_{B+}\rangle + |\alpha_{B-}\rangle$, and the conditioned Wigner functions $W_{\rm REC}$ at the times indicated by the vertical dashed lines in the two trajectories on the left [$\kappa (t_1, t_2)=20, 50$, respectively]. Operating parameters read: $g/\kappa=10$, $B/\kappa=0.5$, and $\gamma/(2\kappa)=0$. The Fock-state basis was truncated at the $50$-th (or $60$-th) photon level, and the JC system was initialized in $|0\rangle |-\rangle$.}
\label{fig:FIG3}
\end{figure*}

The two attractors identified for $\theta=0$ are not placed on equal footing any longer for a finite value of $\theta$. By increasing the LO phase, the first (negative) term in Eq.~\eqref{eq:dE} progressively annihilates the second (positive) term until we operate the homodyne detector with $\cos\theta=1/2$ ($\theta=\pi/3$) or $\cos\theta=-1/2$ ($\theta=2\pi/3$), where the real part of $dE^{\prime}$ vanishes. At the same time, the negative (for $\theta=\pi/3$) or positive (for $\theta=2\pi/3$) imaginary part of $dE$ biases the evolution towards a particular attractor, breaking the state-selection symmetry noted at $\theta=0$. Occasional fluctuations owing to measurement backaction are significantly amplified until they eventually precipitate the switch to the biased attractor, as we can observe in Figs.~\ref{fig:FIG2}(c, d). Certainly, the values $\theta=\pi/3$ and $2\pi/3$ only represent a formal limit; for the majority of individual realizations,  fluctuations $\sim \alpha\, dW$ will induce the irreversible switch for smaller values of $\theta$. 

Upon approaching the ``thermodynamic limit'' $g/\kappa \to \infty$ (entailing $\alpha \to \infty$), any nonzero value of $\theta$ will ``instantly'' collapse the trajectories to only one of the macroscopic attractors $\alpha=\pm g/(2\kappa)$. A notable exception occurs at $\theta=\pi/2$, where $dE^{\prime}$ is real and linear in $\alpha$. For that setting of the LO phase, the trajectory is entirely constrained in the $X=0$ plane of the Bloch sphere for any value of $\alpha$, and the two Stern--Gerlach dressed states are never visited in the evolution. As soon as $\alpha^2 \cos\theta$ (which determines the imaginary part of $dE^{\prime}$ close to $\theta=\pi/2$) takes an appreciable value, however, localization to a particular attractor is restored. 

As a complementary method of record making, we may employ heterodyne instead of homodyne detection, replacing $e^{-i\theta}\,a\,dq_{\theta}(t)$ with $a\,dq(t)$ in Eq.~\eqref{eq:SSE}. Now the complex measured signal (amplitude and phase) takes the form $dq(t)=\sqrt{2\kappa} \langle \psi_{\rm REC}(t)| a^{\dagger}|\psi_{\rm REC}(t) \rangle + dZ$, where $dZ=(dWx + idWy)/\sqrt{2}$ is a complex-valued Wiener increment with covariances $\overline{dZ dZ}=\overline{dZ^{*}dZ^{*}}=0$ and $\overline{dZ^{*}dZ}=dt$. In effect, the LO phase $\theta$ is replaced by a linear function of time with its slope determined by a strong detuning to the output field. We note that heterodyne detection is equivalent to running $x \equiv A_0$ and $y \equiv A_{\pi/2}$ quadrature homodyne measurements in parallel. To accomplish this strategy, we split the signal beam at a $50/50$ beam splitter to measure $x$ on one beam splitter output, and $y$ on the other. Doing so, we will find an equiprobable evolution to one or the other of the distinct dressed states, exactly as it happens for $\theta=0$: effectively, the stability of the $x$-quadrature measurement dominates over the instability of the $y$-quadrature measurement~\cite{Carmichael1999} in the combined dynamics and restores `symmetry' in the alternative outcomes, which is the case in the original Stern--Gerlach experiment.

We now consider drive amplitudes much closer to the bistability threshold, located at $\varepsilon=g^2/(2\kappa)$~\citep{Alsing1991}. We are thus moving to a new regime outside the validity of the secular approximation: the dressed states are self-consistently determined while the effective ``magnetic'' field is no longer a prescribed external field~\citep{Alsing1991, CarmichaelBook2}. Approaching the threshold from above with the LO phase set to $\theta=0$ generates a bimodal distribution in the Bloch-sphere representation, in line with the bistable switching in the homodyne current $i_{0}(t)$ and the bimodal Wigner function $W_{\rm ss}(x+iy)$ of the cavity steady state in Panel I of Fig.~\ref{fig:FIG3}. The photocurrent shows that the metastable states have lifetimes of the order of tens of cavity lifetimes, while $W_{\rm ss}$ exhibits two peaks at the $y$-axis symmetric locations
\begin{equation}\label{eq:AB}
\alpha_{B\pm}\equiv x_{B\pm}+iy_B=A[\pm g/(2\kappa)+iA(\varepsilon/g)]-i\varepsilon/g,
\end{equation} 
where $A\equiv \sqrt{1-[g/(2\varepsilon)]^2(g/\kappa)^2}$. The two peaks are joined by a `skirt' pattern in the phase-space distribution, a sign indicating that we are operating relatively close to the bistability threshold. The `atomic' states are no longer the pair $X=\pm 1$ we previously met, but are instead self-consistently related to the cavity field quadratures~\cite{Alsing1991}. The two distributions in the Bloch sphere, corresponding to the semiclassical amplitudes~\eqref{eq:AB}, are centred about
\begin{equation}
X_{B\pm} + i Y_B= \pm A + i g^2/(2\varepsilon\kappa).
\end{equation}

Setting now $\theta=\pi/2$ conditions a totally different evolution compared to $\theta=0$: not only has the bimodality vanished, but the `atom' trajectory is entirely confined in the $X=0$ plane of the Bloch sphere, as was the case for $\varepsilon/g \gg 1$. An alternative aspect of the dynamics is offered by the conditioned Wigner functions $W_{\rm REC}(x+iy)$ plotted in Fig.~\ref{fig:FIG3}I, corresponding to five ordered instances of time $t_1<t_2<...<t_5$ during the trajectory. At the times $t_2$ and $t_4$, they evince the emergence of the familiar quantum interference pattern associated with a coherent-state superposition of the form $|\alpha_{B+}\rangle + e^{i\phi} |\alpha_{B-}\rangle$~\cite{HarocheBook,Carmichael2013Ch4}, comprising the (un-normalized) conditioned cavity state. The two state amplitudes fluctuate about $\alpha_{B\pm}$, as does the relative phase between the two components. The latter is reflected in the rapidly changing sign of the conditioned Wigner function at the point $(0,y_B)$ located on the axis of symmetry, indicative of the temporal variation of the interference fringes in phase space (see also the phase diffusion discussed in~\cite{Carmichael1999}). Over different realizations, the ensemble average of continuously produced superpositions~\cite{Roy2015} dephases, leaving us with a purely positive Wigner function in the steady state with no trace of the said interference. 

At the times $t_1, t_3$ and $t_5$, the conditioned cavity states track the `skirt' of the steady-state Wigner function, conforming to the periodicity of the interference pattern. The occasional fluctuations ``downwards'' we see in $Y(t)$ of the top-central inset in Fig.~\ref{fig:FIG3}I correspond to deviations from a coherent-state superposition in the cavity state, {\it e.g}, at the time $t_3$.  As the `skirt' disappears and the two peaks separate in phase space with increasing $\varepsilon/g$, fluctuations about the semiclassical value $Y_B=g^2/(2\varepsilon \kappa)$ significantly decrease in the trajectory $Y(t)$; the price to pay is an instability in the system observables and a fading interference pattern in the conditioned bimodal cavity distributions---a suppression of coherence in timescales larger than $\varepsilon^{-1}$ compatible with the validity of the secular approximation. Turning on cavity-inhibited spontaneous emission with $\gamma/(2\kappa) \ll 1$ under direct photodetection of both dissipation channels also continually generates macroscopically distinguishable states. These states, however, have a short lifetime ($\sim \kappa^{-1}$) and are slaved to the phase switch initiated by the atomic collapse that explicitly conditions them~\cite{Carmichael1993QTIV}. In contrast, homodyne detection with $\gamma/(2\kappa) \ll 1$ and $\theta=\pi/2$ stabilizes the coherent-state superpositions in the long timespans where $Y(t)$ doesn't appreciably deviate from $Y_B$. 

The parameter scaling quantum fluctuations for $\gamma/(2\kappa)\to 0$ (the so-called ``zero system size'' limit) coincides with the cavity excitation in the two Stern--Gerlach distinct outcomes under homodyne detection with $\theta=0$, namely $\alpha^2=[g/(2\kappa)]^2$, in the course of individual realizations obtained for $\varepsilon/g \gg 1$. In the ensemble-average sense, it coincides with the asymptotic limit of the steady-state cavity photon number $\langle a^{\dagger}a \rangle_{\rm ss}=[g/(2\kappa)]^2$ obtained from the solution of the ME~\eqref{eq:ME}, while in the mean-field equations of motion it sets the bistability threshold at $\varepsilon/g=g/(2\kappa)$. The latter is confirmed by the emergence of bimodality in the steady-state Wigner functions [Panel II of Fig.~\ref{fig:FIG3}], in two scenarios where the steady-state cavity photon number grows from about $92\%$ to one exceeding $99\%$ of its asymptotic value $[g/(2\kappa)]^2=25$. Individual trajectories, however, reveal a more informative picture of quantum fluctuations than a mere asymptotic scaling.

We remain with the complementary unraveling scheme of homodyne detection where the LO phase is fixed to $\theta=\pi/2$. Panel II of Figure~\ref{fig:FIG3} focuses on the persistence of a stabilized coherent-state superposition with increasing values of $\varepsilon/g$, allowed for by a higher asymptotic excitation amplitude $g/(2\kappa)$ compared to the one selected for Panel I. The conditioned Wigner functions reflect the structure of the ensemble average {\it quasi}probability distribution, complemented by the interference fringes of a coherent-state superposition whose amplitudes are dictated by the steady-state peaks. When we transition from just above threshold in frame (a) to twice the threshold in (b), the time-averaged quadrature amplitude wanes together with the fluctuations in the conditioned cavity photon number, while the frequency of the interference fringes in phase space grows with increasing $\varepsilon/g$. As we mentioned above, a further increase in the drive amplitude gradually induces large fluctuations in the conditioned quadrature phase amplitude $\mathcal{A}_{\pi/2}$ and photon number. Simultaneously, the quantum interference pattern fades out as the ``magnetic'' field of the Stern--Gerlach interaction is slaved to the strong external coherent drive $\varepsilon$. In fact, we can observe in Fig.~\ref{fig:FIG3}II(b) that the side peaks at $\alpha_{B\pm}$ are already relatively higher than the fringe peak in the Wigner function of the superposition state $|\alpha_{B+}\rangle + |\alpha_{B-}\rangle$. This is characteristic of the trend underlying the secular approximation: the side peaks are gradually enhanced over the `fragile' interference fringes. The former remain unstable in time when $\theta$ is set to $\pi/2$. 

As soon as $\theta$ changes to $0$, the two states with amplitudes $\pm g/(2\kappa)$ are stabilized and correlated to the distinct ``spin'' outcomes. The two metastable states identified in the photocurrent $i_0(t)$ in the bottom right inset plot of Fig.~\ref{fig:FIG3}II(b) have longer lifetimes (and larger amplitudes) than their counterparts in Panel I. Their lifetimes diverge for increasing $\varepsilon/g \gg 1$, and they eventually become stationary following the rapid initial random binary state selection in the Stern--Gerlach monitoring scheme. In the Bloch-sphere representation, the two bistable attractors organizing quantum fluctuations move towards the $X$-axis and finally reduce to either fixed point $X=+1$ or $X=-1$, as we saw in Fig.~\ref{fig:FIG2}.

\begin{figure}
\includegraphics[width=0.5\textwidth]{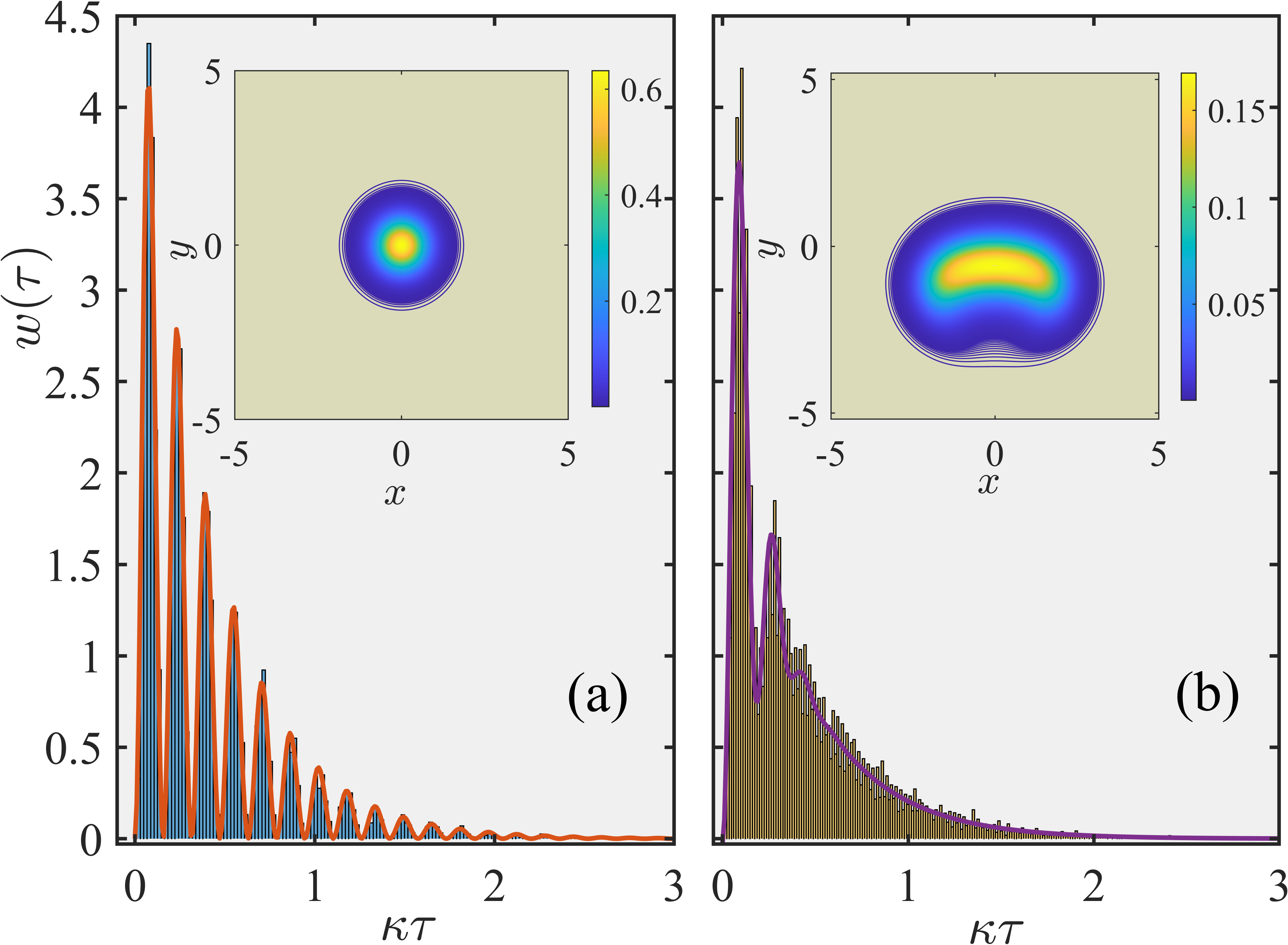}
\caption{{\it Photoelectron statistics in the bad-cavity limit under heterodyne detection.} Histogram of waiting times for the photoelectric emissions to the sides of the cavity (extracted from an individual trajectory along $10^4$ cavity lifetimes) on top of the waiting-time distribution, $w(\tau)$, for $g/\gamma=0.1$ in {\bf (a)} and $1$ in {\bf (b)}. The above exclusive probability density function is calculated using the steady-state solution of the ME~\eqref{eq:ME} and the quantum regression formula. The two insets depict the corresponding steady-state Wigner function $W_{\rm ss} (x+iy)$ for the cavity field. Operating parameters read: $\gamma/\kappa=5$, $\varepsilon/\kappa=20$. In all simulations, the Fock-state basis was truncated at the $20$-th photon level, and the JC system was initialized in $|0\rangle |-\rangle$.}
\label{fig:FIG4}
\end{figure}

In the last part of our discussion, we reinstate cavity-inhibited spontaneous emission. The dissipated radiation is accounted for by `atomic' collapses, $|\psi_{\rm REC}(t)\rangle \to \sigma_{-}|\psi_{\rm REC}(t)\rangle$, at a rate $\gamma \langle\psi_{\rm REC}(t)|\sigma_{+}\sigma_{-}|\psi_{\rm REC}(t)\rangle$~\cite{Reiner2001} interrupting the continuous evolution dictated by the SSE of heterodyne detection, which now reflects the environment encountered by the output field. We focus on the waiting-time distribution of the photoelectrons~\cite{Carmichael1989, Brandes2008} recorded with unit efficiency by the ambient detectors at the sides of the resonator (illustrated in Fig.~\ref{fig:FIG1}). For $g/\gamma \ll 1$, the extracted distribution shows perturbative deviations from the expression of resonance fluorescence, which is proportional to $e^{-\gamma\tau/2} \sin^2(\tau \sqrt{16\varepsilon^2-\gamma^2}/4)$; $\tau$ is the waiting time between photoelectron ``clicks''~\citep{Carmichael1989, CarmichaelBook1}. The corresponding histogram of waiting times (single realization) superposed on the function $w(\tau)$ extracted from the ME and the quantum regression theorem (ensemble average) are depicted in Fig.~\ref{fig:FIG4}(a). They both vanish with a somewhat larger period and attain reduced peaks, yet the salient features follow the exclusive probability density of waiting times in free-space fluorescence. Expectedly, the cavity field state is very close to the vacuum. 

Nonperturbative deviations in the stationary process are observed, however, for $g/\gamma \sim 1$, as squeezed quantum fluctuations in the cavity field occupy an extended region in phase space, with a steady-state intracavity excitation of a couple of photons [see Fig.~\ref{fig:FIG4}(b)]. The formula of $w(\tau)$ from ordinary resonance fluorescence with an effective reduced drive amplitude, $\varepsilon \to \varepsilon - g |\langle a\rangle_{\rm ss}| \approx 0.8 \varepsilon$, as suggested by the form of the JC Hamiltonian~\eqref{eq:JCHam}, is not valid either. For example, the histogram generated from a single trajectory, and the steady-state $w(\tau)$ never reach zero. The latter instance reflects the redistribution of quantum fluctuations driving the induced phased `atomic' dipole. Further Monte Carlo simulations reproduce the degradation of the peak visibility in the trajectory-generated histograms under homodyne detection for any value of $\theta$---the trend is no longer contextual.

In summary, we have explored complementary unravelings of a single ME in distinct regimes motivated by the proposal of a quantum optical analogue to the celebrated Stern--Gerlach experiment. We first deployed balanced homodyne detection~\cite{Eichler2012, CarmichaelBook2} of resonance fluorescence to distinguish between the dressed states of an `atomic' transition under the secular approximation. We proceeded to demonstrate the bias to a particular attractor depending on the LO phase and the intracavity excitation. We thus encountered the issue of quantum contextuality, since the records produced with different LO phases unravel one and the same ME, yet they do not relax to the two dressed states as equiprobable outcomes or may not even produce them whatsoever. Moving now to the occurrence of spontaneous dressed-state polarization, we found a persisting in time phase-diffusing coherent-state superposition~\cite{YurkeStoler1986, CarmichaelSG1994, Carmichael1999} when the LO is set along the direction of the steady-state intracavity amplitude. Remnants of the associated quantum interference pattern are present in the conditioned phase-space distributions in timescales exceeding the lifetimes of the metastable states arising when the LO phase is set to the orthogonal direction. However, the macroscopic state superpositions are prone to structural instability when accessing the operating regime where the Stern--Gerlach apparatus functions in its original conception to produce distinct outcomes. Finally, we moved to a bad-cavity limit~\cite{Savage1988} where no bistability is noted, and deployed the photoelectron waiting-time distribution of the inhibited spontaneous emission photoelectron ``clicks'', recorded in order to assess the squeezed quantum fluctuations of the light fed back to the `atom'.  

The data and codes that support the findings reported herein are openly available on {\it figshare} at the set with this \href{https://doi.org/10.17045/sthlmuni.32099488}{DOI}.

\bibliography{bibliography_SG}

\begin{thebibliography}{47}%
\makeatletter
\providecommand \@ifxundefined [1]{%
 \@ifx{#1\undefined}
}%
\providecommand \@ifnum [1]{%
 \ifnum #1\expandafter \@firstoftwo
 \else \expandafter \@secondoftwo
 \fi
}%
\providecommand \@ifx [1]{%
 \ifx #1\expandafter \@firstoftwo
 \else \expandafter \@secondoftwo
 \fi
}%
\providecommand \natexlab [1]{#1}%
\providecommand \enquote  [1]{``#1''}%
\providecommand \bibnamefont  [1]{#1}%
\providecommand \bibfnamefont [1]{#1}%
\providecommand \citenamefont [1]{#1}%
\providecommand \href@noop [0]{\@secondoftwo}%
\providecommand \href [0]{\begingroup \@sanitize@url \@href}%
\providecommand \@href[1]{\@@startlink{#1}\@@href}%
\providecommand \@@href[1]{\endgroup#1\@@endlink}%
\providecommand \@sanitize@url [0]{\catcode `\\12\catcode `\$12\catcode
  `\&12\catcode `\#12\catcode `\^12\catcode `\_12\catcode `\%12\relax}%
\providecommand \@@startlink[1]{}%
\providecommand \@@endlink[0]{}%
\providecommand \url  [0]{\begingroup\@sanitize@url \@url }%
\providecommand \@url [1]{\endgroup\@href {#1}{\urlprefix }}%
\providecommand \urlprefix  [0]{URL }%
\providecommand \Eprint [0]{\href }%
\providecommand \doibase [0]{https://doi.org/}%
\providecommand \selectlanguage [0]{\@gobble}%
\providecommand \bibinfo  [0]{\@secondoftwo}%
\providecommand \bibfield  [0]{\@secondoftwo}%
\providecommand \translation [1]{[#1]}%
\providecommand \BibitemOpen [0]{}%
\providecommand \bibitemStop [0]{}%
\providecommand \bibitemNoStop [0]{.\EOS\space}%
\providecommand \EOS [0]{\spacefactor3000\relax}%
\providecommand \BibitemShut  [1]{\csname bibitem#1\endcsname}%
\let\auto@bib@innerbib\@empty
\bibitem [{\citenamefont {Sleator}\ \emph {et~al.}(1992)\citenamefont
  {Sleator}, \citenamefont {Pfau}, \citenamefont {Balykin}, \citenamefont
  {Carnal},\ and\ \citenamefont {Mlynek}}]{Sleator1996}%
  \BibitemOpen
  \bibfield  {author} {\bibinfo {author} {\bibfnamefont {T.}~\bibnamefont
  {Sleator}}, \bibinfo {author} {\bibfnamefont {T.}~\bibnamefont {Pfau}},
  \bibinfo {author} {\bibfnamefont {V.}~\bibnamefont {Balykin}}, \bibinfo
  {author} {\bibfnamefont {O.}~\bibnamefont {Carnal}},\ and\ \bibinfo {author}
  {\bibfnamefont {J.}~\bibnamefont {Mlynek}},\ }\bibfield  {title} {\bibinfo
  {title} {Experimental demonstration of the optical stern-gerlach effect},\
  }\href {https://doi.org/10.1103/PhysRevLett.68.1996} {\bibfield  {journal}
  {\bibinfo  {journal} {Phys. Rev. Lett.}\ }\textbf {\bibinfo {volume} {68}},\
  \bibinfo {pages} {1996} (\bibinfo {year} {1992})}\BibitemShut {NoStop}%
\bibitem [{\citenamefont {Blais}\ \emph {et~al.}(2021)\citenamefont {Blais},
  \citenamefont {Grimsmo}, \citenamefont {Girvin},\ and\ \citenamefont
  {Wallraff}}]{Blais2021}%
  \BibitemOpen
  \bibfield  {author} {\bibinfo {author} {\bibfnamefont {A.}~\bibnamefont
  {Blais}}, \bibinfo {author} {\bibfnamefont {A.~L.}\ \bibnamefont {Grimsmo}},
  \bibinfo {author} {\bibfnamefont {S.~M.}\ \bibnamefont {Girvin}},\ and\
  \bibinfo {author} {\bibfnamefont {A.}~\bibnamefont {Wallraff}},\ }\bibfield
  {title} {\bibinfo {title} {Circuit quantum electrodynamics},\ }\href
  {https://doi.org/10.1103/RevModPhys.93.025005} {\bibfield  {journal}
  {\bibinfo  {journal} {Rev. Mod. Phys.}\ }\textbf {\bibinfo {volume} {93}},\
  \bibinfo {pages} {025005} (\bibinfo {year} {2021})}\BibitemShut {NoStop}%
\bibitem [{\citenamefont {García~Ripoll}(2022)}]{GarcíaRipoll2022}%
  \BibitemOpen
  \bibfield  {author} {\bibinfo {author} {\bibfnamefont {J.~J.}\ \bibnamefont
  {García~Ripoll}},\ }\href@noop {} {\emph {\bibinfo {title} {Quantum
  Information and Quantum Optics with Superconducting Circuits}}}\ (\bibinfo
  {publisher} {Cambridge University Press},\ \bibinfo {year}
  {2022})\BibitemShut {NoStop}%
\bibitem [{\citenamefont {Minev}\ \emph {et~al.}(2019)\citenamefont {Minev},
  \citenamefont {Mundhada}, \citenamefont {Shankar}, \citenamefont {Reinhold},
  \citenamefont {Guti{\'e}rrez-J{\'a}uregui}, \citenamefont {Schoelkopf},
  \citenamefont {Mirrahimi}, \citenamefont {Carmichael},\ and\ \citenamefont
  {Devoret}}]{Minev2019}%
  \BibitemOpen
  \bibfield  {author} {\bibinfo {author} {\bibfnamefont {Z.~K.}\ \bibnamefont
  {Minev}}, \bibinfo {author} {\bibfnamefont {S.~O.}\ \bibnamefont {Mundhada}},
  \bibinfo {author} {\bibfnamefont {S.}~\bibnamefont {Shankar}}, \bibinfo
  {author} {\bibfnamefont {P.}~\bibnamefont {Reinhold}}, \bibinfo {author}
  {\bibfnamefont {R.}~\bibnamefont {Guti{\'e}rrez-J{\'a}uregui}}, \bibinfo
  {author} {\bibfnamefont {R.~J.}\ \bibnamefont {Schoelkopf}}, \bibinfo
  {author} {\bibfnamefont {M.}~\bibnamefont {Mirrahimi}}, \bibinfo {author}
  {\bibfnamefont {H.~J.}\ \bibnamefont {Carmichael}},\ and\ \bibinfo {author}
  {\bibfnamefont {M.~H.}\ \bibnamefont {Devoret}},\ }\bibfield  {title}
  {\bibinfo {title} {To catch and reverse a quantum jump mid-flight},\ }\href
  {https://doi.org/10.1038/s41586-019-1287-z} {\bibfield  {journal} {\bibinfo
  {journal} {Nature}\ }\textbf {\bibinfo {volume} {570}},\ \bibinfo {pages}
  {200} (\bibinfo {year} {2019})}\BibitemShut {NoStop}%
\bibitem [{\citenamefont {Murch}\ \emph {et~al.}(2013)\citenamefont {Murch},
  \citenamefont {Weber}, \citenamefont {Macklin},\ and\ \citenamefont
  {Siddiqi}}]{Qubit2013}%
  \BibitemOpen
  \bibfield  {author} {\bibinfo {author} {\bibfnamefont {K.~W.}\ \bibnamefont
  {Murch}}, \bibinfo {author} {\bibfnamefont {S.~J.}\ \bibnamefont {Weber}},
  \bibinfo {author} {\bibfnamefont {C.}~\bibnamefont {Macklin}},\ and\ \bibinfo
  {author} {\bibfnamefont {I.}~\bibnamefont {Siddiqi}},\ }\bibfield  {title}
  {\bibinfo {title} {Observing single quantum trajectories of a superconducting
  quantum bit},\ }\href {https://doi.org/10.1038/nature12539} {\bibfield
  {journal} {\bibinfo  {journal} {Nature}\ }\textbf {\bibinfo {volume} {502}},\
  \bibinfo {pages} {211} (\bibinfo {year} {2013})}\BibitemShut {NoStop}%
\bibitem [{\citenamefont {Weber}\ \emph {et~al.}(2016)\citenamefont {Weber},
  \citenamefont {Murch}, \citenamefont {Kimchi-Schwartz}, \citenamefont
  {Roch},\ and\ \citenamefont {Siddiqi}}]{Weber2016}%
  \BibitemOpen
  \bibfield  {author} {\bibinfo {author} {\bibfnamefont {S.~J.}\ \bibnamefont
  {Weber}}, \bibinfo {author} {\bibfnamefont {K.~W.}\ \bibnamefont {Murch}},
  \bibinfo {author} {\bibfnamefont {M.~E.}\ \bibnamefont {Kimchi-Schwartz}},
  \bibinfo {author} {\bibfnamefont {N.}~\bibnamefont {Roch}},\ and\ \bibinfo
  {author} {\bibfnamefont {I.}~\bibnamefont {Siddiqi}},\ }\bibfield  {title}
  {\bibinfo {title} {Quantum trajectories of superconducting qubits},\ }\href
  {https://doi.org/10.1016/j.crhy.2016.07.007} {\bibfield  {journal} {\bibinfo
  {journal} {Comptes Rendus. Physique}\ }\textbf {\bibinfo {volume} {17}},\
  \bibinfo {pages} {766} (\bibinfo {year} {2016})}\BibitemShut {NoStop}%
\bibitem [{\citenamefont {Gambetta}\ \emph {et~al.}(2008)\citenamefont
  {Gambetta}, \citenamefont {Blais}, \citenamefont {Boissonneault},
  \citenamefont {Houck}, \citenamefont {Schuster},\ and\ \citenamefont
  {Girvin}}]{Gambetta2008}%
  \BibitemOpen
  \bibfield  {author} {\bibinfo {author} {\bibfnamefont {J.}~\bibnamefont
  {Gambetta}}, \bibinfo {author} {\bibfnamefont {A.}~\bibnamefont {Blais}},
  \bibinfo {author} {\bibfnamefont {M.}~\bibnamefont {Boissonneault}}, \bibinfo
  {author} {\bibfnamefont {A.~A.}\ \bibnamefont {Houck}}, \bibinfo {author}
  {\bibfnamefont {D.~I.}\ \bibnamefont {Schuster}},\ and\ \bibinfo {author}
  {\bibfnamefont {S.~M.}\ \bibnamefont {Girvin}},\ }\bibfield  {title}
  {\bibinfo {title} {Quantum trajectory approach to circuit qed: Quantum jumps
  and the zeno effect},\ }\href {https://doi.org/10.1103/PhysRevA.77.012112}
  {\bibfield  {journal} {\bibinfo  {journal} {Phys. Rev. A}\ }\textbf {\bibinfo
  {volume} {77}},\ \bibinfo {pages} {012112} (\bibinfo {year}
  {2008})}\BibitemShut {NoStop}%
\bibitem [{\citenamefont {Breuer}\ and\ \citenamefont
  {Petruccione}(2007)}]{BreuerPetruccione}%
  \BibitemOpen
  \bibfield  {author} {\bibinfo {author} {\bibfnamefont {H.-P.}\ \bibnamefont
  {Breuer}}\ and\ \bibinfo {author} {\bibfnamefont {F.}~\bibnamefont
  {Petruccione}},\ }\href@noop {} {\emph {\bibinfo {title} {The Theory of Open
  Quantum Systems}}}\ (\bibinfo  {publisher} {Oxford University Press},\
  \bibinfo {year} {2007})\BibitemShut {NoStop}%
\bibitem [{\citenamefont {Fink}\ \emph {et~al.}(2008)\citenamefont {Fink},
  \citenamefont {G{\"o}ppl}, \citenamefont {Baur}, \citenamefont {Bianchetti},
  \citenamefont {Leek}, \citenamefont {Blais},\ and\ \citenamefont
  {Wallraff}}]{Fink2008}%
  \BibitemOpen
  \bibfield  {author} {\bibinfo {author} {\bibfnamefont {J.~M.}\ \bibnamefont
  {Fink}}, \bibinfo {author} {\bibfnamefont {M.}~\bibnamefont {G{\"o}ppl}},
  \bibinfo {author} {\bibfnamefont {M.}~\bibnamefont {Baur}}, \bibinfo {author}
  {\bibfnamefont {R.}~\bibnamefont {Bianchetti}}, \bibinfo {author}
  {\bibfnamefont {P.~J.}\ \bibnamefont {Leek}}, \bibinfo {author}
  {\bibfnamefont {A.}~\bibnamefont {Blais}},\ and\ \bibinfo {author}
  {\bibfnamefont {A.}~\bibnamefont {Wallraff}},\ }\bibfield  {title} {\bibinfo
  {title} {Climbing the jaynes--cummings ladder and observing its nonlinearity
  in a cavity qed system},\ }\href {https://doi.org/10.1038/nature07112}
  {\bibfield  {journal} {\bibinfo  {journal} {Nature}\ }\textbf {\bibinfo
  {volume} {454}},\ \bibinfo {pages} {315} (\bibinfo {year}
  {2008})}\BibitemShut {NoStop}%
\bibitem [{\citenamefont {Bishop}\ \emph {et~al.}(2009)\citenamefont {Bishop},
  \citenamefont {Chow}, \citenamefont {Koch}, \citenamefont {Houck},
  \citenamefont {Devoret}, \citenamefont {Thuneberg}, \citenamefont {Girvin},\
  and\ \citenamefont {Schoelkopf}}]{Bishop2009}%
  \BibitemOpen
  \bibfield  {author} {\bibinfo {author} {\bibfnamefont {L.~S.}\ \bibnamefont
  {Bishop}}, \bibinfo {author} {\bibfnamefont {J.~M.}\ \bibnamefont {Chow}},
  \bibinfo {author} {\bibfnamefont {J.}~\bibnamefont {Koch}}, \bibinfo {author}
  {\bibfnamefont {A.~A.}\ \bibnamefont {Houck}}, \bibinfo {author}
  {\bibfnamefont {M.~H.}\ \bibnamefont {Devoret}}, \bibinfo {author}
  {\bibfnamefont {E.}~\bibnamefont {Thuneberg}}, \bibinfo {author}
  {\bibfnamefont {S.~M.}\ \bibnamefont {Girvin}},\ and\ \bibinfo {author}
  {\bibfnamefont {R.~J.}\ \bibnamefont {Schoelkopf}},\ }\bibfield  {title}
  {\bibinfo {title} {Nonlinear response of the vacuum rabi resonance},\ }\href
  {https://doi.org/10.1038/nphys1154} {\bibfield  {journal} {\bibinfo
  {journal} {Nature Physics}\ }\textbf {\bibinfo {volume} {5}},\ \bibinfo
  {pages} {105} (\bibinfo {year} {2009})}\BibitemShut {NoStop}%
\bibitem [{\citenamefont {Kerckhoff}\ \emph {et~al.}(2011)\citenamefont
  {Kerckhoff}, \citenamefont {Armen},\ and\ \citenamefont
  {Mabuchi}}]{Kerckhoff11}%
  \BibitemOpen
  \bibfield  {author} {\bibinfo {author} {\bibfnamefont {J.}~\bibnamefont
  {Kerckhoff}}, \bibinfo {author} {\bibfnamefont {M.~A.}\ \bibnamefont
  {Armen}},\ and\ \bibinfo {author} {\bibfnamefont {H.}~\bibnamefont
  {Mabuchi}},\ }\bibfield  {title} {\bibinfo {title} {Remnants of semiclassical
  bistability in the few-photon regime of cavity qed},\ }\href
  {https://doi.org/10.1364/OE.19.024468} {\bibfield  {journal} {\bibinfo
  {journal} {Opt. Express}\ }\textbf {\bibinfo {volume} {19}},\ \bibinfo
  {pages} {24468} (\bibinfo {year} {2011})}\BibitemShut {NoStop}%
\bibitem [{\citenamefont {Fink}\ \emph {et~al.}(2017)\citenamefont {Fink},
  \citenamefont {Dombi}, \citenamefont {Vukics}, \citenamefont {Wallraff},\
  and\ \citenamefont {Domokos}}]{Fink2017}%
  \BibitemOpen
  \bibfield  {author} {\bibinfo {author} {\bibfnamefont {J.~M.}\ \bibnamefont
  {Fink}}, \bibinfo {author} {\bibfnamefont {A.}~\bibnamefont {Dombi}},
  \bibinfo {author} {\bibfnamefont {A.}~\bibnamefont {Vukics}}, \bibinfo
  {author} {\bibfnamefont {A.}~\bibnamefont {Wallraff}},\ and\ \bibinfo
  {author} {\bibfnamefont {P.}~\bibnamefont {Domokos}},\ }\bibfield  {title}
  {\bibinfo {title} {Observation of the photon-blockade breakdown phase
  transition},\ }\href {https://doi.org/10.1103/PhysRevX.7.011012} {\bibfield
  {journal} {\bibinfo  {journal} {Phys. Rev. X}\ }\textbf {\bibinfo {volume}
  {7}},\ \bibinfo {pages} {011012} (\bibinfo {year} {2017})}\BibitemShut
  {NoStop}%
\bibitem [{\citenamefont {Sett}\ \emph {et~al.}(2024)\citenamefont {Sett},
  \citenamefont {Hassani}, \citenamefont {Phan}, \citenamefont {Barzanjeh},
  \citenamefont {Vukics},\ and\ \citenamefont {Fink}}]{Sett2024}%
  \BibitemOpen
  \bibfield  {author} {\bibinfo {author} {\bibfnamefont {R.}~\bibnamefont
  {Sett}}, \bibinfo {author} {\bibfnamefont {F.}~\bibnamefont {Hassani}},
  \bibinfo {author} {\bibfnamefont {D.}~\bibnamefont {Phan}}, \bibinfo {author}
  {\bibfnamefont {S.}~\bibnamefont {Barzanjeh}}, \bibinfo {author}
  {\bibfnamefont {A.}~\bibnamefont {Vukics}},\ and\ \bibinfo {author}
  {\bibfnamefont {J.~M.}\ \bibnamefont {Fink}},\ }\bibfield  {title} {\bibinfo
  {title} {Emergent macroscopic bistability induced by a single superconducting
  qubit},\ }\href {https://doi.org/10.1103/PRXQuantum.5.010327} {\bibfield
  {journal} {\bibinfo  {journal} {PRX Quantum}\ }\textbf {\bibinfo {volume}
  {5}},\ \bibinfo {pages} {010327} (\bibinfo {year} {2024})}\BibitemShut
  {NoStop}%
\bibitem [{\citenamefont {Carmichael}\ \emph {et~al.}(1994)\citenamefont
  {Carmichael}, \citenamefont {Kochan},\ and\ \citenamefont
  {Tian}}]{CarmichaelSG1994}%
  \BibitemOpen
  \bibfield  {author} {\bibinfo {author} {\bibfnamefont {H.~J.}\ \bibnamefont
  {Carmichael}}, \bibinfo {author} {\bibfnamefont {P.}~\bibnamefont {Kochan}},\
  and\ \bibinfo {author} {\bibfnamefont {L.}~\bibnamefont {Tian}},\ }\bibinfo
  {title} {Coherent states and open quantum systems: A comment on the
  stern-gerlach experiment and schr\"{o}dinger's cat},\ in\ \href
  {https://doi.org/10.1142/9789814503839_0006} {\emph {\bibinfo {booktitle}
  {Coherent States}}}\ (\bibinfo {year} {1994})\ pp.\ \bibinfo {pages}
  {75--91}\BibitemShut {NoStop}%
\bibitem [{\citenamefont {Carmichael}(1999{\natexlab{a}})}]{Carmichael1999}%
  \BibitemOpen
  \bibfield  {author} {\bibinfo {author} {\bibfnamefont {H.~J.}\ \bibnamefont
  {Carmichael}},\ }\bibfield  {title} {\bibinfo {title} {Quantum jumps
  revisited: An overview of quantum trajectory theory},\ }in\ \href@noop {}
  {\emph {\bibinfo {booktitle} {Quantum Future From Volta and Como to the
  Present and Beyond}}},\ \bibinfo {editor} {edited by\ \bibinfo {editor}
  {\bibfnamefont {P.}~\bibnamefont {Blanchard}}\ and\ \bibinfo {editor}
  {\bibfnamefont {A.}~\bibnamefont {Jadczyk}}}\ (\bibinfo  {publisher}
  {Springer Berlin Heidelberg},\ \bibinfo {address} {Berlin, Heidelberg},\
  \bibinfo {year} {1999})\ pp.\ \bibinfo {pages} {15--36}\BibitemShut {NoStop}%
\bibitem [{\citenamefont {Carmichael}(2008)}]{CarmichaelBook2}%
  \BibitemOpen
  \bibfield  {author} {\bibinfo {author} {\bibfnamefont {H.}~\bibnamefont
  {Carmichael}},\ }\href@noop {} {\emph {\bibinfo {title} {Statistical Methods
  in Quantum Optics 2}}}\ (\bibinfo  {publisher} {Springer, Berlin, Germany},\
  \bibinfo {year} {2008})\ Chap.\ \bibinfo {chapter} {16, 18}\BibitemShut
  {NoStop}%
\bibitem [{Car(1993{\natexlab{a}})}]{Carmichael1993QTIII}%
  \BibitemOpen
  \bibinfo {title} {{Quantum Trajectories III}},\ in\ \href
  {https://doi.org/10.1007/978-3-540-47620-7_10} {\emph {\bibinfo {booktitle}
  {An Open Systems Approach to Quantum Optics}}}\ (\bibinfo  {publisher}
  {Springer Berlin Heidelberg},\ \bibinfo {address} {Berlin, Heidelberg},\
  \bibinfo {year} {1993})\ pp.\ \bibinfo {pages} {140--154},\ \bibinfo {note}
  {{L}ectures Presented at the Universit{\'e} Libre de Bruxelles October 28 to
  November 4, 1991}\BibitemShut {NoStop}%
\bibitem [{\citenamefont {Wiseman}\ and\ \citenamefont
  {Milburn}(1993)}]{Wiseman1993}%
  \BibitemOpen
  \bibfield  {author} {\bibinfo {author} {\bibfnamefont {H.~M.}\ \bibnamefont
  {Wiseman}}\ and\ \bibinfo {author} {\bibfnamefont {G.~J.}\ \bibnamefont
  {Milburn}},\ }\bibfield  {title} {\bibinfo {title} {Quantum theory of
  field-quadrature measurements},\ }\href
  {https://doi.org/10.1103/PhysRevA.47.642} {\bibfield  {journal} {\bibinfo
  {journal} {Phys. Rev. A}\ }\textbf {\bibinfo {volume} {47}},\ \bibinfo
  {pages} {642} (\bibinfo {year} {1993})}\BibitemShut {NoStop}%
\bibitem [{\citenamefont {Glauber}(1963{\natexlab{a}})}]{Glauber1963}%
  \BibitemOpen
  \bibfield  {author} {\bibinfo {author} {\bibfnamefont {R.~J.}\ \bibnamefont
  {Glauber}},\ }\bibfield  {title} {\bibinfo {title} {Photon correlations},\
  }\href {https://doi.org/10.1103/PhysRevLett.10.84} {\bibfield  {journal}
  {\bibinfo  {journal} {Phys. Rev. Lett.}\ }\textbf {\bibinfo {volume} {10}},\
  \bibinfo {pages} {84} (\bibinfo {year} {1963}{\natexlab{a}})}\BibitemShut
  {NoStop}%
\bibitem [{\citenamefont {Glauber}(1963{\natexlab{b}})}]{Glauber1963II}%
  \BibitemOpen
  \bibfield  {author} {\bibinfo {author} {\bibfnamefont {R.~J.}\ \bibnamefont
  {Glauber}},\ }\bibfield  {title} {\bibinfo {title} {The quantum theory of
  optical coherence},\ }\href {https://doi.org/10.1103/PhysRev.130.2529}
  {\bibfield  {journal} {\bibinfo  {journal} {Phys. Rev.}\ }\textbf {\bibinfo
  {volume} {130}},\ \bibinfo {pages} {2529} (\bibinfo {year}
  {1963}{\natexlab{b}})}\BibitemShut {NoStop}%
\bibitem [{\citenamefont {Kelley}\ and\ \citenamefont
  {Kleiner}(1964)}]{KelleyKleiner1964}%
  \BibitemOpen
  \bibfield  {author} {\bibinfo {author} {\bibfnamefont {P.~L.}\ \bibnamefont
  {Kelley}}\ and\ \bibinfo {author} {\bibfnamefont {W.~H.}\ \bibnamefont
  {Kleiner}},\ }\bibfield  {title} {\bibinfo {title} {Theory of electromagnetic
  field measurement and photoelectron counting},\ }\href
  {https://doi.org/10.1103/PhysRev.136.A316} {\bibfield  {journal} {\bibinfo
  {journal} {Phys. Rev.}\ }\textbf {\bibinfo {volume} {136}},\ \bibinfo {pages}
  {A316} (\bibinfo {year} {1964})}\BibitemShut {NoStop}%
\bibitem [{\citenamefont {Saleh}(1978)}]{Saleh1978}%
  \BibitemOpen
  \bibfield  {author} {\bibinfo {author} {\bibfnamefont {B.}~\bibnamefont
  {Saleh}},\ }\bibinfo {title} {Photoelectron events: A doubly stochastic
  poisson process or theory of photoelectron statistics},\ in\ \href
  {https://doi.org/10.1007/978-3-540-37311-7_5} {\emph {\bibinfo {booktitle}
  {Photoelectron Statistics: With Applications to Spectroscopy and Optical
  Communication}}}\ (\bibinfo  {publisher} {Springer Berlin Heidelberg},\
  \bibinfo {address} {Berlin, Heidelberg},\ \bibinfo {year} {1978})\ pp.\
  \bibinfo {pages} {160--280}\BibitemShut {NoStop}%
\bibitem [{Car(1993{\natexlab{b}})}]{Carmichael1993QTI}%
  \BibitemOpen
  \bibinfo {title} {{Master Equations and Sources I}},\ in\ \href
  {https://doi.org/10.1007/978-3-540-47620-7_2} {\emph {\bibinfo {booktitle}
  {An Open Systems Approach to Quantum Optics}}}\ (\bibinfo  {publisher}
  {Springer Berlin Heidelberg},\ \bibinfo {address} {Berlin, Heidelberg},\
  \bibinfo {year} {1993})\ pp.\ \bibinfo {pages} {5--21},\ \bibinfo {note}
  {{L}ectures Presented at the Universit{\'e} Libre de Bruxelles October 28 to
  November 4, 1991}\BibitemShut {NoStop}%
\bibitem [{\citenamefont {Carmichael}\ \emph {et~al.}(1989)\citenamefont
  {Carmichael}, \citenamefont {Singh}, \citenamefont {Vyas},\ and\
  \citenamefont {Rice}}]{Carmichael1989}%
  \BibitemOpen
  \bibfield  {author} {\bibinfo {author} {\bibfnamefont {H.~J.}\ \bibnamefont
  {Carmichael}}, \bibinfo {author} {\bibfnamefont {S.}~\bibnamefont {Singh}},
  \bibinfo {author} {\bibfnamefont {R.}~\bibnamefont {Vyas}},\ and\ \bibinfo
  {author} {\bibfnamefont {P.~R.}\ \bibnamefont {Rice}},\ }\bibfield  {title}
  {\bibinfo {title} {Photoelectron waiting times and atomic state reduction in
  resonance fluorescence},\ }\href {https://doi.org/10.1103/PhysRevA.39.1200}
  {\bibfield  {journal} {\bibinfo  {journal} {Phys. Rev. A}\ }\textbf {\bibinfo
  {volume} {39}},\ \bibinfo {pages} {1200} (\bibinfo {year}
  {1989})}\BibitemShut {NoStop}%
\bibitem [{\citenamefont {Alsing}\ and\ \citenamefont
  {Carmichael}(1991)}]{Alsing1991}%
  \BibitemOpen
  \bibfield  {author} {\bibinfo {author} {\bibfnamefont {P.}~\bibnamefont
  {Alsing}}\ and\ \bibinfo {author} {\bibfnamefont {H.~J.}\ \bibnamefont
  {Carmichael}},\ }\bibfield  {title} {\bibinfo {title} {Spontaneous
  dressed-state polarization of a coupled atom and cavity mode},\ }\href
  {https://doi.org/10.1088/0954-8998/3/1/003} {\bibfield  {journal} {\bibinfo
  {journal} {Quantum Optics: Journal of the European Optical Society Part B}\
  }\textbf {\bibinfo {volume} {3}},\ \bibinfo {pages} {13} (\bibinfo {year}
  {1991})}\BibitemShut {NoStop}%
\bibitem [{\citenamefont {Barchielli}\ and\ \citenamefont
  {Belavkin}(1991)}]{Barchielli1991}%
  \BibitemOpen
  \bibfield  {author} {\bibinfo {author} {\bibfnamefont {A.}~\bibnamefont
  {Barchielli}}\ and\ \bibinfo {author} {\bibfnamefont {V.~P.}\ \bibnamefont
  {Belavkin}},\ }\bibfield  {title} {\bibinfo {title} {Measurements continuous
  in time and a posteriori states in quantum mechanics},\ }\href
  {https://doi.org/10.1088/0305-4470/24/7/022} {\bibfield  {journal} {\bibinfo
  {journal} {Journal of Physics A: Mathematical and General}\ }\textbf
  {\bibinfo {volume} {24}},\ \bibinfo {pages} {1495} (\bibinfo {year}
  {1991})}\BibitemShut {NoStop}%
\bibitem [{\citenamefont {Dalibard}\ \emph {et~al.}(1992)\citenamefont
  {Dalibard}, \citenamefont {Castin},\ and\ \citenamefont
  {M\o{}lmer}}]{Dalibard1992}%
  \BibitemOpen
  \bibfield  {author} {\bibinfo {author} {\bibfnamefont {J.}~\bibnamefont
  {Dalibard}}, \bibinfo {author} {\bibfnamefont {Y.}~\bibnamefont {Castin}},\
  and\ \bibinfo {author} {\bibfnamefont {K.}~\bibnamefont {M\o{}lmer}},\
  }\bibfield  {title} {\bibinfo {title} {Wave-function approach to dissipative
  processes in quantum optics},\ }\href
  {https://doi.org/10.1103/PhysRevLett.68.580} {\bibfield  {journal} {\bibinfo
  {journal} {Phys. Rev. Lett.}\ }\textbf {\bibinfo {volume} {68}},\ \bibinfo
  {pages} {580} (\bibinfo {year} {1992})}\BibitemShut {NoStop}%
\bibitem [{\citenamefont {Mølmer}\ and\ \citenamefont
  {Castin}(1996)}]{Molmer1996}%
  \BibitemOpen
  \bibfield  {author} {\bibinfo {author} {\bibfnamefont {K.}~\bibnamefont
  {Mølmer}}\ and\ \bibinfo {author} {\bibfnamefont {Y.}~\bibnamefont
  {Castin}},\ }\bibfield  {title} {\bibinfo {title} {Monte carlo wavefunctions
  in quantum optics},\ }\href {https://doi.org/10.1088/1355-5111/8/1/007}
  {\bibfield  {journal} {\bibinfo  {journal} {Quantum and Semiclassical Optics:
  Journal of the European Optical Society Part B}\ }\textbf {\bibinfo {volume}
  {8}},\ \bibinfo {pages} {49} (\bibinfo {year} {1996})}\BibitemShut {NoStop}%
\bibitem [{\citenamefont {Carmichael}(2013)}]{Carmichael2013Ch4}%
  \BibitemOpen
  \bibfield  {author} {\bibinfo {author} {\bibfnamefont {H.~J.}\ \bibnamefont
  {Carmichael}},\ }\bibinfo {title} {Quantum open systems},\ in\ \href
  {https://doi.org/10.1142/9789814460354_0004} {\emph {\bibinfo {booktitle}
  {Strong Light-Matter Coupling: From Atoms to Solid-State Systems}}}\
  (\bibinfo {year} {2013})\ Chap.~\bibinfo {chapter} {4}, pp.\ \bibinfo {pages}
  {99--153}\BibitemShut {NoStop}%
\bibitem [{\citenamefont {Hatridge}\ \emph {et~al.}(2013)\citenamefont
  {Hatridge}, \citenamefont {Shankar}, \citenamefont {Mirrahimi}, \citenamefont
  {Schackert}, \citenamefont {Geerlings}, \citenamefont {Brecht}, \citenamefont
  {Sliwa}, \citenamefont {Abdo}, \citenamefont {Frunzio}, \citenamefont
  {Girvin}, \citenamefont {Schoelkopf},\ and\ \citenamefont
  {Devoret}}]{Hatridge2013}%
  \BibitemOpen
  \bibfield  {author} {\bibinfo {author} {\bibfnamefont {M.}~\bibnamefont
  {Hatridge}}, \bibinfo {author} {\bibfnamefont {S.}~\bibnamefont {Shankar}},
  \bibinfo {author} {\bibfnamefont {M.}~\bibnamefont {Mirrahimi}}, \bibinfo
  {author} {\bibfnamefont {F.}~\bibnamefont {Schackert}}, \bibinfo {author}
  {\bibfnamefont {K.}~\bibnamefont {Geerlings}}, \bibinfo {author}
  {\bibfnamefont {T.}~\bibnamefont {Brecht}}, \bibinfo {author} {\bibfnamefont
  {K.~M.}\ \bibnamefont {Sliwa}}, \bibinfo {author} {\bibfnamefont
  {B.}~\bibnamefont {Abdo}}, \bibinfo {author} {\bibfnamefont {L.}~\bibnamefont
  {Frunzio}}, \bibinfo {author} {\bibfnamefont {S.~M.}\ \bibnamefont {Girvin}},
  \bibinfo {author} {\bibfnamefont {R.~J.}\ \bibnamefont {Schoelkopf}},\ and\
  \bibinfo {author} {\bibfnamefont {M.~H.}\ \bibnamefont {Devoret}},\
  }\bibfield  {title} {\bibinfo {title} {Quantum back-action of an individual
  variable-strength measurement},\ }\href
  {https://doi.org/10.1126/science.1226897} {\bibfield  {journal} {\bibinfo
  {journal} {Science}\ }\textbf {\bibinfo {volume} {339}},\ \bibinfo {pages}
  {178} (\bibinfo {year} {2013})}\BibitemShut {NoStop}%
\bibitem [{\citenamefont {Savage}\ and\ \citenamefont
  {Carmichael}(1988)}]{Savage1988}%
  \BibitemOpen
  \bibfield  {author} {\bibinfo {author} {\bibfnamefont {C.}~\bibnamefont
  {Savage}}\ and\ \bibinfo {author} {\bibfnamefont {H.}~\bibnamefont
  {Carmichael}},\ }\bibfield  {title} {\bibinfo {title} {Single atom optical
  bistability},\ }\href {https://doi.org/10.1109/3.7075} {\bibfield  {journal}
  {\bibinfo  {journal} {IEEE Journal of Quantum Electronics}\ }\textbf
  {\bibinfo {volume} {24}},\ \bibinfo {pages} {1495} (\bibinfo {year}
  {1988})}\BibitemShut {NoStop}%
\bibitem [{\citenamefont {Kilin}\ and\ \citenamefont
  {Krinitskaya}(1991)}]{Kilin91}%
  \BibitemOpen
  \bibfield  {author} {\bibinfo {author} {\bibfnamefont {S.~Y.}\ \bibnamefont
  {Kilin}}\ and\ \bibinfo {author} {\bibfnamefont {T.~B.}\ \bibnamefont
  {Krinitskaya}},\ }\bibfield  {title} {\bibinfo {title} {Single-atom phase
  bistability in a fundamental model of quantum optics},\ }\href
  {https://doi.org/10.1364/JOSAB.8.002289} {\bibfield  {journal} {\bibinfo
  {journal} {J. Opt. Soc. Am. B}\ }\textbf {\bibinfo {volume} {8}},\ \bibinfo
  {pages} {2289} (\bibinfo {year} {1991})}\BibitemShut {NoStop}%
\bibitem [{\citenamefont {Alsing}\ \emph {et~al.}(1992)\citenamefont {Alsing},
  \citenamefont {Guo},\ and\ \citenamefont {Carmichael}}]{Alsing1992}%
  \BibitemOpen
  \bibfield  {author} {\bibinfo {author} {\bibfnamefont {P.}~\bibnamefont
  {Alsing}}, \bibinfo {author} {\bibfnamefont {D.-S.}\ \bibnamefont {Guo}},\
  and\ \bibinfo {author} {\bibfnamefont {H.~J.}\ \bibnamefont {Carmichael}},\
  }\bibfield  {title} {\bibinfo {title} {Dynamic stark effect for the
  jaynes-cummings system},\ }\href {https://doi.org/10.1103/PhysRevA.45.5135}
  {\bibfield  {journal} {\bibinfo  {journal} {Phys. Rev. A}\ }\textbf {\bibinfo
  {volume} {45}},\ \bibinfo {pages} {5135} (\bibinfo {year}
  {1992})}\BibitemShut {NoStop}%
\bibitem [{\citenamefont {Armen}(2009)}]{Armen2009}%
  \BibitemOpen
  \bibfield  {author} {\bibinfo {author} {\bibfnamefont {M.~A.}\ \bibnamefont
  {Armen}},\ }\href
  {https://resolver.caltech.edu/CaltechETD:etd-05262009-100436} {\bibinfo
  {title} {Bifurcations in single atom cavity {QED}}} (\bibinfo {year}
  {2009}),\ \bibinfo {note} {dissertation (Ph.D.)}\BibitemShut {NoStop}%
\bibitem [{\citenamefont {Carmichael}(2015)}]{Carmichael2015}%
  \BibitemOpen
  \bibfield  {author} {\bibinfo {author} {\bibfnamefont {H.~J.}\ \bibnamefont
  {Carmichael}},\ }\bibfield  {title} {\bibinfo {title} {Breakdown of photon
  blockade: A dissipative quantum phase transition in zero dimensions},\ }\href
  {https://doi.org/10.1103/PhysRevX.5.031028} {\bibfield  {journal} {\bibinfo
  {journal} {Phys. Rev. X}\ }\textbf {\bibinfo {volume} {5}},\ \bibinfo {pages}
  {031028} (\bibinfo {year} {2015})}\BibitemShut {NoStop}%
\bibitem [{\citenamefont {Curtis}\ \emph {et~al.}(2021)\citenamefont {Curtis},
  \citenamefont {Boettcher}, \citenamefont {Young}, \citenamefont {Maghrebi},
  \citenamefont {Carmichael}, \citenamefont {Gorshkov},\ and\ \citenamefont
  {Foss-Feig}}]{Curtis2021}%
  \BibitemOpen
  \bibfield  {author} {\bibinfo {author} {\bibfnamefont {J.~B.}\ \bibnamefont
  {Curtis}}, \bibinfo {author} {\bibfnamefont {I.}~\bibnamefont {Boettcher}},
  \bibinfo {author} {\bibfnamefont {J.~T.}\ \bibnamefont {Young}}, \bibinfo
  {author} {\bibfnamefont {M.~F.}\ \bibnamefont {Maghrebi}}, \bibinfo {author}
  {\bibfnamefont {H.}~\bibnamefont {Carmichael}}, \bibinfo {author}
  {\bibfnamefont {A.~V.}\ \bibnamefont {Gorshkov}},\ and\ \bibinfo {author}
  {\bibfnamefont {M.}~\bibnamefont {Foss-Feig}},\ }\bibfield  {title} {\bibinfo
  {title} {Critical theory for the breakdown of photon blockade},\ }\href
  {https://doi.org/10.1103/PhysRevResearch.3.023062} {\bibfield  {journal}
  {\bibinfo  {journal} {Phys. Rev. Res.}\ }\textbf {\bibinfo {volume} {3}},\
  \bibinfo {pages} {023062} (\bibinfo {year} {2021})}\BibitemShut {NoStop}%
\bibitem [{\citenamefont {Reiner}\ \emph {et~al.}(2001)\citenamefont {Reiner},
  \citenamefont {Smith}, \citenamefont {Orozco}, \citenamefont {Carmichael},\
  and\ \citenamefont {Rice}}]{Reiner2001}%
  \BibitemOpen
  \bibfield  {author} {\bibinfo {author} {\bibfnamefont {J.~E.}\ \bibnamefont
  {Reiner}}, \bibinfo {author} {\bibfnamefont {W.~P.}\ \bibnamefont {Smith}},
  \bibinfo {author} {\bibfnamefont {L.~A.}\ \bibnamefont {Orozco}}, \bibinfo
  {author} {\bibfnamefont {H.~J.}\ \bibnamefont {Carmichael}},\ and\ \bibinfo
  {author} {\bibfnamefont {P.~R.}\ \bibnamefont {Rice}},\ }\bibfield  {title}
  {\bibinfo {title} {Time evolution and squeezing of the field amplitude in
  cavity qed},\ }\href {https://doi.org/10.1364/JOSAB.18.001911} {\bibfield
  {journal} {\bibinfo  {journal} {J. Opt. Soc. Am. B}\ }\textbf {\bibinfo
  {volume} {18}},\ \bibinfo {pages} {1911} (\bibinfo {year}
  {2001})}\BibitemShut {NoStop}%
\bibitem [{\citenamefont {Tan}(1999)}]{Tan1999}%
  \BibitemOpen
  \bibfield  {author} {\bibinfo {author} {\bibfnamefont {S.~M.}\ \bibnamefont
  {Tan}},\ }\bibfield  {title} {\bibinfo {title} {A computational toolbox for
  quantum and atomic optics},\ }\href
  {https://doi.org/10.1088/1464-4266/1/4/312} {\bibfield  {journal} {\bibinfo
  {journal} {Journal of Optics B: Quantum and Semiclassical Optics}\ }\textbf
  {\bibinfo {volume} {1}},\ \bibinfo {pages} {424} (\bibinfo {year}
  {1999})}\BibitemShut {NoStop}%
\bibitem [{\citenamefont {Cohen-Tannoudji}\ and\ \citenamefont
  {Reynaud}(1977)}]{CohenTannoudji1977}%
  \BibitemOpen
  \bibfield  {author} {\bibinfo {author} {\bibfnamefont {C.}~\bibnamefont
  {Cohen-Tannoudji}}\ and\ \bibinfo {author} {\bibfnamefont {S.}~\bibnamefont
  {Reynaud}},\ }\bibfield  {title} {\bibinfo {title} {Dressed-atom description
  of resonance fluorescence and absorption spectra of a multi-level atom in an
  intense laser beam},\ }\href {https://doi.org/10.1088/0022-3700/10/3/005}
  {\bibfield  {journal} {\bibinfo  {journal} {Journal of Physics B: Atomic and
  Molecular Physics}\ }\textbf {\bibinfo {volume} {10}},\ \bibinfo {pages}
  {345} (\bibinfo {year} {1977})}\BibitemShut {NoStop}%
\bibitem [{\citenamefont {Schleich}(2001)}]{SchleichCh3}%
  \BibitemOpen
  \bibfield  {author} {\bibinfo {author} {\bibfnamefont {W.~P.}\ \bibnamefont
  {Schleich}},\ }\bibinfo {title} {Wigner function},\ in\ \href
  {https://doi.org/https://doi.org/10.1002/3527602976.ch3} {\emph {\bibinfo
  {booktitle} {Quantum Optics in Phase Space}}}\ (\bibinfo  {publisher} {John
  Wiley \& Sons, Ltd},\ \bibinfo {year} {2001})\ Chap.~\bibinfo {chapter} {3},
  pp.\ \bibinfo {pages} {67--98}\BibitemShut {NoStop}%
\bibitem [{\citenamefont {Carmichael}(1999{\natexlab{b}})}]{CarmichaelBook1}%
  \BibitemOpen
  \bibfield  {author} {\bibinfo {author} {\bibfnamefont {H.}~\bibnamefont
  {Carmichael}},\ }\href@noop {} {\emph {\bibinfo {title} {Statistical Methods
  in Quantum Optics 1, Master Equations and Fokker-Planck Equations}}}\
  (\bibinfo  {publisher} {Springer, Berlin, Germany},\ \bibinfo {year} {1999})\
  Chap.\ \bibinfo {chapter} {1, 2, 3, 4}\BibitemShut {NoStop}%
\bibitem [{\citenamefont {Haroche}\ and\ \citenamefont
  {Raimond}(2006)}]{HarocheBook}%
  \BibitemOpen
  \bibfield  {author} {\bibinfo {author} {\bibfnamefont {S.}~\bibnamefont
  {Haroche}}\ and\ \bibinfo {author} {\bibfnamefont {J.-M.}\ \bibnamefont
  {Raimond}},\ }\href
  {https://doi.org/10.1093/acprof:oso/9780198509141.001.0001} {\emph {\bibinfo
  {title} {{Exploring the Quantum: Atoms, Cavities, and Photons}}}}\ (\bibinfo
  {publisher} {Oxford University Press},\ \bibinfo {year} {2006})\BibitemShut
  {NoStop}%
\bibitem [{\citenamefont {Roy}\ \emph {et~al.}(2015)\citenamefont {Roy},
  \citenamefont {Leghtas}, \citenamefont {Stone}, \citenamefont {Devoret},\
  and\ \citenamefont {Mirrahimi}}]{Roy2015}%
  \BibitemOpen
  \bibfield  {author} {\bibinfo {author} {\bibfnamefont {A.}~\bibnamefont
  {Roy}}, \bibinfo {author} {\bibfnamefont {Z.}~\bibnamefont {Leghtas}},
  \bibinfo {author} {\bibfnamefont {A.~D.}\ \bibnamefont {Stone}}, \bibinfo
  {author} {\bibfnamefont {M.}~\bibnamefont {Devoret}},\ and\ \bibinfo {author}
  {\bibfnamefont {M.}~\bibnamefont {Mirrahimi}},\ }\bibfield  {title} {\bibinfo
  {title} {Continuous generation and stabilization of mesoscopic field
  superposition states in a quantum circuit},\ }\href
  {https://doi.org/10.1103/PhysRevA.91.013810} {\bibfield  {journal} {\bibinfo
  {journal} {Phys. Rev. A}\ }\textbf {\bibinfo {volume} {91}},\ \bibinfo
  {pages} {013810} (\bibinfo {year} {2015})}\BibitemShut {NoStop}%
\bibitem [{Car(1993{\natexlab{c}})}]{Carmichael1993QTIV}%
  \BibitemOpen
  \bibinfo {title} {{Quantum Trajectories IV}},\ in\ \href
  {https://doi.org/10.1007/978-3-540-47620-7_11} {\emph {\bibinfo {booktitle}
  {An Open Systems Approach to Quantum Optics}}}\ (\bibinfo  {publisher}
  {Springer Berlin Heidelberg},\ \bibinfo {address} {Berlin, Heidelberg},\
  \bibinfo {year} {1993})\ pp.\ \bibinfo {pages} {155--173},\ \bibinfo {note}
  {{L}ectures Presented at the Universit{\'e} Libre de Bruxelles October 28 to
  November 4, 1991}\BibitemShut {NoStop}%
\bibitem [{\citenamefont {Brandes}(2008)}]{Brandes2008}%
  \BibitemOpen
  \bibfield  {author} {\bibinfo {author} {\bibfnamefont {T.}~\bibnamefont
  {Brandes}},\ }\bibfield  {title} {\bibinfo {title} {Waiting times and noise
  in single particle transport},\ }\href
  {https://doi.org/https://doi.org/10.1002/andp.20085200707} {\bibfield
  {journal} {\bibinfo  {journal} {Annalen der Physik}\ }\textbf {\bibinfo
  {volume} {520}},\ \bibinfo {pages} {477} (\bibinfo {year}
  {2008})}\BibitemShut {NoStop}%
\bibitem [{\citenamefont {Eichler}\ \emph {et~al.}(2012)\citenamefont
  {Eichler}, \citenamefont {Bozyigit},\ and\ \citenamefont
  {Wallraff}}]{Eichler2012}%
  \BibitemOpen
  \bibfield  {author} {\bibinfo {author} {\bibfnamefont {C.}~\bibnamefont
  {Eichler}}, \bibinfo {author} {\bibfnamefont {D.}~\bibnamefont {Bozyigit}},\
  and\ \bibinfo {author} {\bibfnamefont {A.}~\bibnamefont {Wallraff}},\
  }\bibfield  {title} {\bibinfo {title} {Characterizing quantum microwave
  radiation and its entanglement with superconducting qubits using linear
  detectors},\ }\href {https://doi.org/10.1103/PhysRevA.86.032106} {\bibfield
  {journal} {\bibinfo  {journal} {Phys. Rev. A}\ }\textbf {\bibinfo {volume}
  {86}},\ \bibinfo {pages} {032106} (\bibinfo {year} {2012})}\BibitemShut
  {NoStop}%
\bibitem [{\citenamefont {Yurke}\ and\ \citenamefont
  {Stoler}(1986)}]{YurkeStoler1986}%
  \BibitemOpen
  \bibfield  {author} {\bibinfo {author} {\bibfnamefont {B.}~\bibnamefont
  {Yurke}}\ and\ \bibinfo {author} {\bibfnamefont {D.}~\bibnamefont {Stoler}},\
  }\bibfield  {title} {\bibinfo {title} {Generating quantum mechanical
  superpositions of macroscopically distinguishable states via amplitude
  dispersion},\ }\href {https://doi.org/10.1103/PhysRevLett.57.13} {\bibfield
  {journal} {\bibinfo  {journal} {Phys. Rev. Lett.}\ }\textbf {\bibinfo
  {volume} {57}},\ \bibinfo {pages} {13} (\bibinfo {year} {1986})}\BibitemShut
  {NoStop}%
\end{thebibliography}%

\end{document}